\documentstyle[aps,epsf,twocolumn]{revtex} 
\begin{document}
\title
{
Casimir effect in dielectrics: Bulk Energy Contribution
}
\author{
C. E. Carlson$^{*}$, C. Molina--Par\'{\i}s$^{+}$,
J. P\'erez--Mercader$^{++}$, and Matt Visser$^{+++}$
}
\address{
$^{*}$Physics Department, College of William and Mary,
Williamsburg, Virginia 23187\\
$^{+}$Theoretical Division, Los Alamos National Laboratory,
Los Alamos, New Mexico 87545\\
$^{++}$Laboratorio de Astrof\'{\i}sica Espacial y F\'{\i}sica
Fundamental, Apartado 50727, 28080 Madrid\\
$^{+++}$Physics Department, Washington University,
St. Louis, Missouri 63130-4899\\
}
\date{hepth/9702007; 31 January 1997}
\maketitle
\begin{table*}[tb!]
\begin{center}
\begin{minipage}{7.0 in}
\parshape=1 0.75in 5.5in 
\indent 
{\small 
In a recent series of papers, Schwinger discussed a process that 
he called the Dynamical Casimir Effect. The key essence of this
effect is the {\em change} in zero-point energy associated with
any {\em change} in a dielectric medium. (In particular, if the
change in the dielectric medium is taken to be the growth or collapse
of a bubble, this effect {\em may} have relevance to sonoluminescence.)
The kernel of Schwinger's result is that the change in Casimir
energy is proportional to the change in volume of the dielectric,
plus finite-volume corrections. Other papers have called into
question this result, claiming that the volume term should actually
be discarded, and that the dominant term remaining is proportional
to the surface area of the dielectric.  In this communication,
which is an expansion of an earlier letter on the same topic, we
present a careful and critical review of the relevant analyses.
We find that the Casimir energy, defined as the change in zero-point
energy due to a change in the medium, has at leading order a bulk
volume dependence. This is in {\em full} agreement with Schwinger's
result, once the correct physical question is asked.  We have
nothing new to say about sonoluminescence itself.
}
\end{minipage}
\end{center}
\end{table*}
\clearpage
\def\Re{{\rm Re}}
\def\Im{{\rm Im}}
\section{Schwinger's calculation}

Several years ago, the late Julian Schwinger wrote a series of
papers~\cite{Schwinger0,Schwinger1,Schwinger2} wherein he calculated
the Casimir energy released in the collapse of a spherically
symmetric bubble or cavity. (For the evolution of Schwinger's ideas
on this topic, see~\cite{Schwinger3,Schwinger4,Schwinger5,Schwinger6}.)
Using source theory, he derived in a simple and elegant way a
formula for the energy release involved in the collapse. He found
from a straightforward application of the action principle that
(for each polarization state) the {\em ``dielectric energy, relative
to the zero energy of the vacuum, [is given] by}

\begin{equation}
E = - V \int \frac{d^3\vec{k}}{(2 \pi)^3} 
\frac{1}{2} [\hbar c] k  
\left( 1 - \frac{1}{\sqrt{\epsilon}} \right).
\label{E-Schwinger-0}
\end{equation}

\noindent
{\em So the Casimir energy of a uniform dielectric is negative''}. 
This was then applied by Schwinger as a model for sonoluminescence.

{From} this, and following Schwinger, we conclude that a dielectric
slab of material with a spherical vacuum cavity of radius $R$ has
a {\em higher} Casimir energy than the slab of material with the
vacuum cavity re--filled with material of the same dielectric
constant.  This energy (per polarization state) can be computed by
introducing an ultraviolet momentum cut-off $K$ into the previous
expression for the energy~\footnote{For generic dielectrics, this
wave-number cut-off is related to the high-wave-number asymptotic
behavior of the dispersion relation: it is the scale at which the
dielectric dispersion relation begins to approach the vacuum
dispersion relation. For the particular case of sonoluminescence,
this wave-number cut-off---deduced from the dielectric properties
of the medium---implies a minimum wavelength for the electromagnetic
radiation emitted in the collapse of the bubble.}

\begin{eqnarray}
E_{\small cavity}
&=&  +\left[ V \int \frac{d^3\vec{k}}{(2 \pi)^3}
\frac{1}{2} \hbar c k \left( 1 - \frac{1}{\sqrt{\epsilon}} \right)
\right] +\dots
\nonumber\\
&=&+\frac{4\pi}{3}R^3 \; \int_0^K dk \frac{1}{2}
{4 \pi k^3 \over(2\pi)^3} \hbar c
 \left(1 - \frac{1}{\sqrt{\epsilon}}
\right) +\dots
\nonumber\\
&=&\frac{1}{12 \pi} \hbar c R^3 K^4  
\left(1 - \frac{1}{\sqrt{\epsilon}} \right) +\dots
\end{eqnarray}

\noindent
In general, this volume contribution will be the {\em dominant}
contribution.\footnote{The dots denote finite-volume corrections.
We shall investigate the leading finite-volume term more fully in
a separate publication~\cite{CMPV-states}.}

In view of the elegance and simplicity of this result, it is natural
to ask whether it can also be derived by more traditional quantum
field theoretic means.  Indeed, the existence of such a volume
contribution is easy to verify on general physical grounds:

\noindent
(1) For instance, we know that the effective action in 3+1 dimensions
contains divergences which range from quartic to logarithmic, in
addition to finite contributions.  Furthermore (as is well known),
the ``cosmological constant'' contribution (the quartic divergence)
{\em never} vanishes unless the theory has very special
symmetries~\cite{Weinberg} (as for example, in the case of
supersymmetric theories). Thus energy densities that go as ({\rm
cut-off})${}^4$ are {\em generic} in (3+1) dimensions.

\noindent 
(2) The cut-off dependence of the energy should not be considered
alarming: Dielectrics are condensed matter systems, and as such
abound in physically defined and physically meaningful
cut-offs---everything from the plasma frequency to the interatomic
spacing may be considered as a candidate for the cut-off. The
interesting physics comes in deciding exactly which cut-off is
physically relevant in the current situation.

\noindent
(3) We can view Schwinger's result in elementary terms as simply
the difference in zero-point-energies obtained by integrating the
difference in photon dispersion relations over the density of
states

\begin{equation}
E_{\small cavity} = + 2 V \int \frac{d^3\vec{k}}{(2 \pi)^3} \frac{1}{2}
\hbar  \left[ c k - \omega(k)  \right] + \cdots
\end{equation}

\noindent
The dots again denote finite-volume corrections. The energy above
is multiplied by $2$ because unlike eq. $(\ref{E-Schwinger-0})$ we
have already included the contributions from the two polarization
states.  To actually calculate this energy difference we require
a suitable physical model for $\omega(k)$.

\noindent
(4) Alternatively, we could perform an explicit quantum field
theoretic calculation of the Casimir energy in some specific model
problem and thereby verify Schwinger's result. We have performed
such a calculation, and report the results in this paper. 

We find full agreement with the calculation and results of
Schwinger~\cite{Schwinger0},  and disagreement with some subsequent
papers. For instance, the calculation presented in
references~\cite{Milton80,Milton95,Milton96} computes the Casimir
energy associated with a spherical ball of radius $R$, dielectric
constant $\epsilon_1$, and permeability $\mu_1$, embedded in an
infinite medium with dielectric constant $\epsilon_2$ and permeability
$\mu_2$.

That calculation ultimately asserts that the dominant contribution
is not the volume term but a surface correction\footnote{Actually
there is some ambiguity even on this point --- some, but not all,
of the comments in those papers could be interpreted as suggesting
that the surface term should also be discarded, and that the dominant
term is actually of order $\hbar c/R$. This is tantamount to
arbitrarily setting to zero anything that depends on the cut-off
and simply relying on naive dimensional analysis to guess the form
of the answer. For physical dielectrics with physical cut-offs this
suggestion is not tenable.}  which is proportional to $R^2 K^3
(\epsilon_1 -\epsilon_2)^2$.  We, however, have re-analyzed these
calculations and find that the volume term is in fact present and
{\it {dominant}} except for very small bubbles.  (The present paper
is a more detailed and complete presentation of a result previously
announced in~\cite{CMPV-short}.)

We have re-analyzed this calculation in several different  and
complementary ways:
\\
(A) We have extended one method of doing the calculation, namely
the method of doing a mode sum over the difference of zero point
energies, to further illuminate the physics behind the cut-off by
introducing a number of simple general and robust arguments. We
show that generically surface terms do in fact show up as {\em
sub-dominant} corrections to the dominant volume effect. The general
arguments contained in this section of the paper are particularly
useful in that we can deal with {\em arbitrary} shapes and not be
limited by requirements of spherical symmetry. This material is
reported in section~\ref{modesum}.
\\
(B)
We have analyzed previous calculations to see where they differ
from our own. We first clarify the physical content of the different
configurations whose zero point energies are to be subtracted from
each other to get the Casimir energy.  The results can be compactly
represented as regulated integrals involving Green functions.  Some
prior calculations misapplied the subtraction scheme required to
get rid of the ``vacuum'' contribution, and we can isolate the
difference in energy between these prior calculations and the
correct one. This energy difference can be calculated by an elementary
application of general quantum field theory results and we thus
regain Schwinger's volume term. This result is again general and
not limited by requirements of spherical symmetry.  The qualitative
parts of this analysis, including a general demonstration of the
existence of the volume term, is in section~\ref{pictures}, and
numerical coefficients of the volume term are obtained by this
shape independent method in section~\ref{green}.
\\
(C)
For the case of a spherical dielectric ball,  one can also address
the problem by using expansions of Green functions in spherical
coordinates, following the formalism of~\cite{Milton80,Milton95,Milton96}.
We focus first on the volume term of the Casimir energy, {\em i.e.},
on the energy difference between some prior calculations and the
full calculation.  Given our previous discussion,  if the volume
term is considered in isolation the use of specific coordinates
makes the calculation more complicated than it has to be.  However,
we wish to see the same answer emerge using this method. The energy
difference is given as a sum over a series of integrals involving
Ricatti--Bessel functions which we evaluate {\em exactly} using
generalized addition theorems for spherical Bessel functions. The
result, presented in section~\ref{spherical}, is in complete
agreement with the quantum field theory calculation and our extensions
of the mode sum calculation, and also in complete agreement with
Schwinger's original claim~\cite{Schwinger0}.
\\
(D)
For completeness, we also present the full calculation using the
general solutions to the electromagnetic fields developed in
references~\cite{Milton80,Milton95,Milton96} using the dyadic
formalism.  We compute the energy difference between (Case I) an
otherwise uniform medium with dielectric constant $\epsilon_2$ and
permeability $\mu_2$, but with a spherical cavity of radius $R$
containing dielectric constant $\epsilon_1$ and permeability $\mu_1$,
and (Case II) a completely uniform medium with dielectric constant
$\epsilon_2$ and permeability $\mu_2$.  This energy difference is
{\em again} given as a sum over a series of integrals involving
Ricatti--Bessel functions. Some of these sums can be evaluated
explicitly while others can only be evaluated by using an asymptotic
analysis of the type presented in~\cite{Milton80,Milton95,Milton96}
. We verify the existence of both volume and sub-dominant surface
contributions, and present these results in section~\ref{full}.

\section{Mode sum over zero point energies}
\label{modesum}

We wish here to present a simple derivation of Schwinger's 
result and to make some extensions of, and comments about, it.

\subsection{Dielectric embedded in vacuum}

Ultimately, the physics underlying the Casimir effect emerges from
the fact that every eigenmode of the photon field has zero point
energy $E_n = (1/2)\hbar \omega_n$. The Casimir energy is simply
the difference in zero point energies between any two well  defined
physical situations

\begin{equation}
E_{\small Casimir}(A|B) = 
\sum_n {1\over2} \hbar \left[\omega_n(A) - \omega_n(B)\right].
\end{equation}

\noindent
Sometimes we will need a regulator to make sense of this energy
difference, though in many other cases of physical interest (as in
dielectrics) the physics of the problem will automatically regulate
the difference for us and make the results finite. But adding over
all eigen-modes is prohibitively difficult, so it is in general
more productive to replace the sum over states by an integral over
the density of states,

\begin{equation}
\sum_n \sim V \int { d^3 \vec k\over (2\pi)^3} + \cdots
\end{equation}

Suppose we have a finite volume $V$ of some bulk dielectric in
which the dispersion relation for photons is given by some function
$\omega(k)$, which describes the photon frequency as a function of
the wave number (three-momentum) $k$. To keep infra-red divergences
under control introduce a regulator by putting the whole universe
in a box of finite volume $V_\infty$. Then to calculate the total
zero-point energy of the electromagnetic field we simply have to
sum the photon energies over all momenta (and polarizations), using
the usual and elementary density of states: $\hbox{[Volume]} \,
d^3 \vec k /(2\pi)^3$.  (In the next section we shall look at
finite-volume corrections to this density of states.)

Including photon modes both inside and outside the dielectric 
body is achieved by calculating

\begin{eqnarray}
E_{\small dielectric} &=& 
2 V \int {d^3 \vec k\over(2\pi)^3} \; {1\over2} \hbar \;
\omega(k)
\nonumber\\
&+& 2 (V_\infty - V) \int {d^3 \vec k\over(2\pi)^3} \; 
{1\over2} \hbar \; c k.
\label{E-dielectric}
\end{eqnarray}

\noindent
Note that outside the dielectric we have simply taken the photon
dispersion relation to be that of the Minkowski vacuum $\omega (k)=
c k$.

At low frequencies, we know that the dispersion relation for
a dielectric is simply summarized by the zero-frequency refractive
index $n$. That is

\begin{equation}
\omega(k) \to c k/n \qquad {\rm as} \qquad k \to 0.
\end{equation}

\noindent
On the other hand, at high enough frequencies, the photons propagate
freely through the dielectric: They are then simply free photons
traveling through the empty vacuum between individual atoms. Thus

\begin{equation}
\omega(k) \to c k \qquad {\rm as} \qquad k \to \infty.
\end{equation}

In the absence of the dielectric, we can calculate the total
zero-point energy of the Minkowski vacuum as

\begin{equation}
E_{\small vacuum} = 
2 V_\infty \int {d^3 \vec k\over(2\pi)^3} \; {1\over2} \hbar \; c k.
\label{E-vacuum}
\end{equation}

\noindent
Let us now subtract the two  zero-point energies given in 
eqns. $(\ref{E-dielectric})$ and $(\ref{E-vacuum})$ as

\begin{eqnarray}
E_{\small Casimir} 
&\equiv&  E_{\small dielectric} - E_{\small vacuum}
\nonumber \\
&=&
2 V \int {d^3 \vec k\over(2\pi)^3} \; 
{1\over2} \hbar \; [\omega(k) - c k],
\end{eqnarray}

\noindent 
which defines $E_{\small Casimir}$ for a dielectric body 
embedded in vacuum.

{From} the above asymptotic analysis, we know that the integrand
must go to zero at large wave-number.  {\em In fact, for any real
physical dielectric the integrand must go to zero sufficiently
rapidly to make the integral converge since, after all, we are
talking about a real physical difference in energies.} To actually
calculate this energy difference we require a suitable physical
model for $\omega(k)$.

\begin{figure}[htb]
\vbox{\hfil\epsfbox{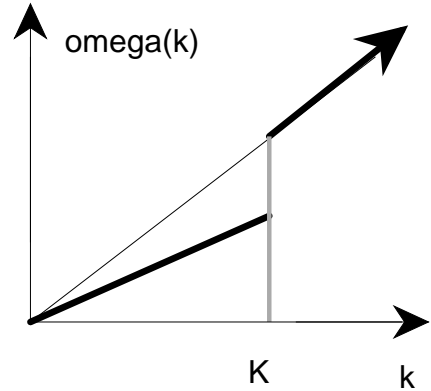}\hfil}
\caption
{\label{F-model-1}
Schwinger's model dispersion relation.}
\end{figure}

Schwinger's calculation reported in Ref.~\cite{Schwinger0} is
equivalent to choosing the particularly simple model

\begin{equation}
\omega(k) = {c k\over n} \; \Theta(K-k) + ck \; \Theta(k-K)
\end{equation}

\noindent
whose physical meaning is immediately seen from the presence of
the Heaviside step function $\Theta(x)$; in addition, $K$ is  a
wave-number (three-momentum)  which characterizes the transition
from dielectric-like behavior to vacuum-like behavior. (See figure
\ref{F-model-1}.) For a dielectric body embedded in vacuum this
particular model gives

\begin{equation}
E_{\small Casimir} = 
{1\over8\pi^2} V \; \hbar c \; K^4 \left[ {1\over  n} - 1 \right].
\label{E-dielectric-in-vacuum}
\end{equation}


\noindent
Note that the cut-off $K$ describes real physics: It is a surrogate
for all the complicated physics that would be required to make
a detailed model for the dielectric to vacuum transition in the 
dispersion relation.

Of course, this can all be recast in terms of a wave-number dependent
refractive index

\begin{equation}
{1\over n(k)} \equiv {\omega(k)\over c \; k},
\end{equation}

\noindent
and then, an integration by parts yields

\begin{eqnarray}
E_{\small Casimir}  
&=&
{ V \hbar \over 6 \pi^2 } \int d(k^3)\; [\omega(k) - ck]
\nonumber\\
&=&
{ V \hbar \over 6 \pi^2 } 
\Bigg[ \left.\left(k^3\; 
     [\omega(k) - ck]\right)\right|_0^\infty  
\nonumber\\
&& \qquad -  \int [d\omega(k) - d(ck)] k^3 \Bigg].
\end{eqnarray}

The boundary term vanishes because of the asymptotic behavior 
of $\omega(k)$, and the two substitutions $k = \omega(k) n/c$ and 
$k=\omega/c$ then give 

\begin{equation}
E_{\small Casimir} = + {V \hbar \over 6 \pi^2 c^3} 
\int_0^\infty \omega^3 [1-n^3(\omega)] d\omega.
\end{equation}

\begin{figure}[htb]
\vbox{\hfil\epsfbox{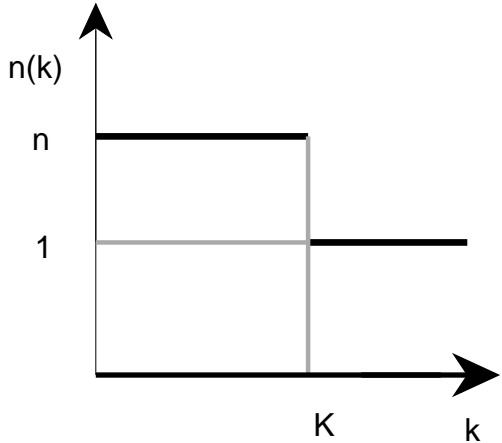}\hfil}
\caption
{\label{F-model-2}
Schwinger's model for the refractive index.
Refractive index as a function of wave-number.}
\end{figure}

\noindent
In terms of the refractive index, of course, Schwinger's model
for the dispersion relation becomes (see figure \ref{F-model-2})

\begin{equation}
n(k) = n \; \Theta(K-k) + \Theta(k-K),
\end{equation}


\begin{figure}[htb]
\vbox{\hfil\epsfbox{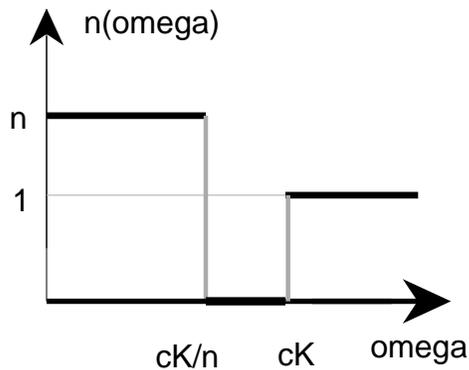}\hfil}
\caption
{\label{F-model-3}
Schwinger's model for the refractive index.
Refractive index as a function of frequency.}
\end{figure}

\noindent
and equivalently (see figure \ref{F-model-3})

\begin{equation}
n(\omega) = n \; \Theta([Kc/n]-\omega) + \Theta(\omega-Kc).
\end{equation}

We now go on to the more general case of two dielectric media.


\subsection{Two dielectric media}

Suppose we now have two dielectrics to consider. We will deal with
the situation where a dielectric body with dispersion relation
$\omega_1(k)$ is embedded in an infinite slab of dielectric with
dispersion relation $\omega_2(k)$. The total zero-point energy is
easily written down

\begin{eqnarray}
E_{\small embedded-body} 
&=& 
2 V \int {d^3 \vec k\over(2\pi)^3} \; {1\over2} \hbar \; \omega_1(k)
\nonumber\\
&+& 
2 (V_\infty - V) \int {d^3 \vec k\over(2\pi)^3} \; 
{1\over2} \hbar \; \omega_2(k).
\end{eqnarray}

If the embedded body is removed, and the hole simply filled-in with
the embedding medium, we can simply calculate the {\em{new}} total
zero-point energy as

\begin{equation}
E_{\small homogeneous} 
= 
2 V_\infty \int {d^3 \vec k\over(2\pi)^3} \; 
{1\over2} \hbar \; \omega_2(k).
\end{equation}

\noindent
We {\em define} the Casimir energy by subtracting these two zero-point
energies

\begin{eqnarray}
E_{\small Casimir} &\equiv&  
E_{\small embedded-body} - E_{\small homogeneous}
\nonumber \\
&=&
2 V \int {d^3 \vec  k\over(2\pi)^3} \; 
{1\over2} \hbar \; [\omega_1(k) - \omega_2(k)].
\label{E-two-dielectrics}
\end{eqnarray}

\noindent
The physical import of this definition is clear: The Casimir energy
is defined as the {\em change} in the zero-point energy due to a
{\em change} in the medium.

{From} the general considerations in the previous subsection we
know that the integrand must (still) go to zero at large wave-number,
and in fact, for any pair of real physical dielectrics the integrand
must go to zero sufficiently rapidly to make the integral converge.

The integration by parts discussed above can now be extended to
yield

\begin{equation}
E_{\small Casimir} = + {V \hbar \over 6 \pi^2 c^3} 
\int_0^\infty \omega^3 [n_2^3(\omega)-n_1^3(\omega)] d\omega.
\end{equation}

\noindent
While the difference between the refractive indices in the above
expression goes to zero sufficiently rapidly to make the integral
converge, it must be noted that the prefactor of $\omega^3$ implies
that the net Casimir energy will be relatively sensitive to the
high frequency behavior of the refractive indices.

If we now use Schwinger's simple wave-number cut-off, we find that

\begin{equation}
E_{\small Casimir} = 
{1\over8\pi^2} V \; \hbar c \; 
\left\{ K_1^4 \left[ {1\over  n_1} - 1 \right] -  
        K_2^4 \left[ {1\over  n_2} - 1 \right] \right\}.
\label{E-two-cut-offs}
\end{equation}

\noindent
Notice that depending on the precise values of the two refractive
indices and the two cut-offs, we can easily change the {\em sign}
of the Casimir energy.

In particular, if we have a cavity containing vacuum embedded in an
otherwise uniform dielectric  $n_1=1$, and can for simplicity 
take $n_2=n$ and $K_2=K$, in which case

\begin{equation}
E_{\small Casimir} = 
-{1\over8\pi^2} V \; \hbar c \; K^4 \left[ {1\over  n} - 1 \right].
\label{E-cavity-in-dielectric}
\end{equation}

\noindent
Note that this is exactly the {\em negative} of the result for a
dielectric body embedded in vacuum. [See Eq.
(\ref{E-dielectric-in-vacuum}).] This observation serves to drive
home the fact that the Casimir energy makes no sense until one has
carefully specified the {\em two} physical situations whose energy
{\em difference} is being calculated.

\subsection{A general cut-off}

In order to appreciate the degree to which Schwinger's expression is
generic, notice that the dispersion relation can {\em always} be
written in the form

\begin{equation}
\omega(k) = c k + ck \left[ {1\over n} - 1 \right] {\cal F}(k/K).
\end{equation}

\begin{figure}[htb]
\vbox{\hfil\epsfbox{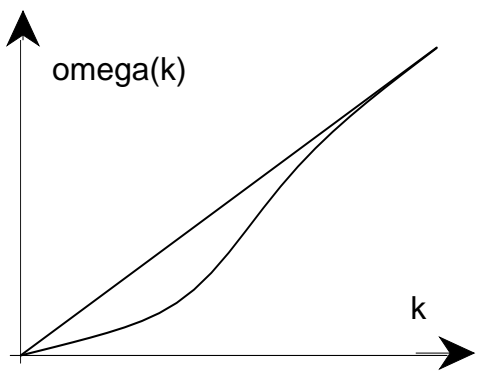}\hfil}
\caption
{\label{F-model-4}
Generic model dispersion relation.}
\end{figure}

\noindent
This expression serves to {\em define} the function ${\cal F}(k/K)$,
and the cut-off scale $K$. We know from the general discussion that
${\cal F}(0) = 1$, while ${\cal F}(\infty) = 0$.  It is convenient
to normalize $\cal F$ by ${\cal F}(1) = 1/2$, and thus implicitly
fix $K$ by the requirement that $\omega(K) = c K (n+1)/(2n)$.  With
this notation $K$ is the wave-number at which $\omega(k)/k$ has
fallen half-way from its low-momentum dielectric value to its high
wave-number free-space value. (See figure \ref{F-model-4}.)


\noindent
In terms of the wave-number dependent refractive index,

\begin{equation}
{1\over n(k) } = 1 +
\left[ {1\over n} - 1 \right] {\cal F}(k/K).
\end{equation}

\noindent
The Casimir energy difference is now simply written as

\begin{equation}
E_{\small Casimir} = {1\over2\pi^2} V \; \hbar c \; K^4 
\left[ {1\over  n} -1 \right]
\int_0^\infty x^3 {\cal F}(x) dx.
\end{equation}

\indent
This extension of Schwinger's result makes manifest 
the salient points of the Casimir energy in dielectrics:
\\
{}(1) The (dominant) contribution to the Casimir zero-point-energy
is a bulk effect proportional to the volume of the {\it displaced}
dielectric,
\\
{}(2) and equivalently, the bulk Casimir energy {\em density} is
\begin{equation}
\rho = {1\over2\pi^2} \hbar c \; K^4 \left[ {1\over n} -1 \right]
 \int_0^\infty x^3 {\cal F}(x) dx.
\end{equation}
\\
{}(3) The dependence on refractive index at low momentum is explicitly known.
\\
{}(4) There is an explicit ultraviolet cut-off $K$, and the Casimir
energy is proportional to this cut-off to the fourth power---{\em
this cut-off is real physics, and is not an artifact to be renormalized
away.}
\\
{}(5) It is only an accident of history that many, but by no means
all, of the early Casimir effect calculations were for configurations
where the cut-off dependence happens to vanish. (See~\cite{BVW}.)
\\
{}(6) There is an overall dimensionless constant that arises from
the detailed behavior of the dispersion relation as a function of
wave-number---this remaining overall factor cannot be calculated
without developing a specific detailed model for the dispersion.
\\
{}(7) The key physical insight underlying the Casimir effect is
that the presence of a dielectric or conductor {\em changes} the
photon dispersion relation and thereby {\em changes} the total
zero-point energy.

If we  wish to tackle the problem of one dielectric embedded in
another, the relevant generalizations are clear: we simply
write down individual cut-off functions ${\cal F}_{(1,2)}$ and cut-off
scales $K_{(1,2)}$ for each dielectric.  The Casimir energy
difference is now given by

\begin{eqnarray}
E_{\small Casimir} &=& {1\over2\pi^2} V \; \hbar c \; 
\Bigg\{ 
K_1^4 \left[ {1\over  n_1} -1 \right] 
\int_0^\infty x^3 {\cal F}_1(x) dx 
\nonumber\\
&& \qquad\qquad -
K_2^4 \left[ {1\over  n_2} -1 \right] 
\int_0^\infty x^3 {\cal F}_2(x) dx  
\Bigg\}.
\nonumber\\ 
\end{eqnarray} 

Next we turn to a discussion of the finite-volume effects.

\subsection{Finite-volume effects}

We now look at contributions arising from sub-dominant finite-volume
corrections to the density of states. The key observation here is that
the existence of finite-volume corrections proportional to the
surface area of the dielectric is a {\em generic} result.  The fact
that Milton {\em et al.}\ encountered a surface-tension term
proportional to $(\hbox{cut-off})^3$ is hereby explained on general
physical grounds without recourse to special function theory.

Now, the fact that the dominant contribution to the Casimir energy
is proportional only to volume is of course, merely a reflection of the
fact that the canonical bulk expression for the density of states
is proportional to volume: $\hbox{[Volume]}\, d^3 \vec k/(2\pi)^3$.
It is reasonably well-known, though perhaps not so elementary, that
the density of states is in general modified by finite volume
effects. Thus in general we should write

\begin{equation}
\sum_n \sim V \int { d^3 \vec k\over (2\pi)^3} +
            S \int \xi { d^3 \vec k \over (2\pi)^3 k } + 
	    \cdots
\end{equation}

\noindent
These are the first two terms in an asymptotic expansion in $1/k$.
A discussion of the general existence of such terms can be found
in the textbook by Pathria~\cite{Pathria}, while a more extensive
discussion has been given in the paper by Balian and
Bloch~\cite{Balian-Bloch}.  For Dirichlet, von Neumann, and Robin
boundary conditions the dimensionless variable $\xi$ is a known
function of the boundary conditions imposed. For dielectric junction
conditions, the case of interest in the current problem, the
situation is considerably more complicated and will be dealt with
in a forthcoming paper. For now we just point out that, introducing
separate quantities $\xi_{\small in}$ and $\xi_{\small out}$ for
the density of states inside and outside the dielectric body, the
first finite-volume contribution to the Casimir energy is of the
form

\begin{equation}
S \int { d^3 \vec k \over (2\pi)^3 k } \hbar 
\left[ 
\xi_{\small in} \; \omega_{\small in}(k) + 
\xi_{\small out} \; \omega_{\small out}(k) 
\right],
\end{equation}

\noindent
that is to say

\begin{equation}
S \int { d^3 \vec k \over (2\pi)^3 } \hbar c 
\left[ 
{\xi_{\small in}\over n_{\small in}(k)} + 
{\xi_{\small out}\over n_{\small out}(k)} 
\right].  
\end{equation}

\noindent
For a simple wave-number cut-off, \'a la Schwinger, this becomes

\begin{equation}
{1\over6\pi^2} S  K^3  \hbar c 
\left[ 
{\xi_{\small in}\over n_{\small in}} + 
{\xi_{\small out}\over n_{\small out}} 
\right],
\end{equation}

\noindent
which turns out to be in full agreement with the results enunciated
in~\cite{Milton80,Milton95,Milton96}. To actually complete the
calculation we need to evaluate both $\xi_{\small in}$ and $\xi_{\small
out}$ as functions of refractive index for both TE and TM modes.
This will be the subject of a forthcoming paper~\cite{CMPV-states}.

\subsection{A potential source of confusion}

A potential source of confusion should now be noted before it leads
to trouble: If we introduce a naive cut-off in frequency (energy) rather
than wave-number (three-momentum) then the Casimir zero-point energy
appears to have a different behavior as a function of refractive
index.  This is merely a reflection of the fact that frequency
cut-offs are in general ill-behaved, commonly leading to multi-branched
dispersion relations.

For definiteness, consider what at first sight would seem to be the
frequency cut-off version of Schwinger's model

\begin{equation}
k(\omega) = {n \omega\over c} \; \Theta(\omega_0-\omega) +
            {\omega \over c}  \; \Theta(\omega- \omega_0).
\end{equation}

\begin{figure}[htb]
\vbox{\hfil\epsfbox{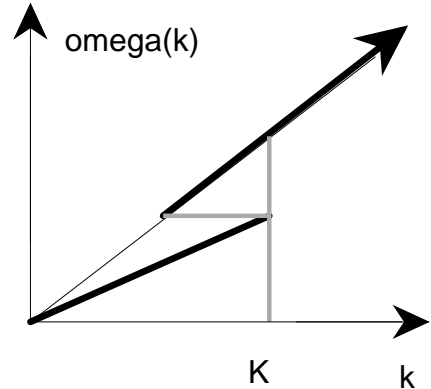}\hfil}
\caption
{\label{F-model-5}
A naive frequency cut-off for the model dispersion relation: Note
the unphysical double-valued nature of the dispersion.}
\end{figure}

\noindent
Here $\Theta(x)$ is again the Heaviside step function, while
$\omega_0$ is now a frequency  which characterizes the transition
from dielectric-like behavior to vacuum-like behavior.  (See
figure \ref{F-model-3}.) If we try to invert this function to obtain
the dispersion relation $\omega(k)$, {\em we see that it 
is double-valued in the range $k\in[\omega_0/c,n \omega_0/c]$}.
Momenta in this range will be doubly-counted, and lead to 
unphysical results. (See figure \ref{F-model-5}.)


For this particular model, we get

\begin{equation}
E_{\small Casimir} = 
{1\over8\pi^2} V \hbar c \left({\omega_0\over c}\right)^4 
\left[ n^3 - 1 \right].
\end{equation}

\noindent
Note that the double-counting has been sufficient to change the
sign of the bulk Casimir energy! (See eq. (12)). Physically, the
sign as determined in Schwinger's calculation is the correct one
as can be seen from a  general argument: Suppose only that the
dispersion relation is single-valued and that $\omega(k) < ck$,
(equivalently $n(k) > 1$), then the sign as determined by Schwinger
is correct.

The $[n^3-1]$ behavior encountered above is in fact generic for
frequency-based or energy-based cut-offs.  In particular, (as we
shall see below) if we impose point-slitting in time as our regulator,
and then define energy differences by subtracting two time-split
energies which are time-split at the same physical time, then we
are effectively imposing a frequency-based regulator and will obtain
the $[n^3-1]$ behavior. On the other hand, if we time-split using
``optical time'', the parameter $\tau_{\small optical} = \tau_{\small
physical}/n$, then we are effectively imposing a wave-number-based
regulator, and we generically get a $[(1/n)-1]$ behavior.

A wave-number based regulator, following Schwinger's prescription,
is by far more physical than a naive frequency-based regulator.
The only reason we belabor this point is because many calculations
are carried out using naive time-splitting in physical time, and
should be slightly modified (so that we time-split in optical time)
before being compared to Schwinger's result. Otherwise one is led
into meaningless results.

We emphasize that this is a minor side issue that does not affect
issues of volume dependence versus surface dependence, and thus
does not alter the cut-off dependence. At worst these technical
issues influence the behavior as a function of refractive index,
and even then the {\em order} ($O[n-1]$) of the Casimir energy for
dilute media ($n\approx1$) will not be affected.

\section{Physical description of the calculations and 
existence of the volume term}
\label{pictures}

The calculations presented by Milton {\em et
al.}~\cite{Milton80,Milton95,Milton96} avoid the density-of-states
point of view by explicitly calculating the eigen-modes (equivalently,
Green functions) for a specific model configuration: a dielectric
ball (of dielectric constant $\epsilon_1$ and permeability $\mu_1$)
embedded in an infinite slab of (different) dielectric constant
$\epsilon_2$ and permeability $\mu_2$.  They then explicitly
integrated over  these eigen-modes to calculate the Casimir
energy.

Note that an important limitation of any such calculation is that
while the density-of-states argument applies to dielectrics of
arbitrary shape, any attempt at explicitly calculating eigen-modes
must be restricted to systems of extremely high symmetry---such as for
example half-spaces, slabs, or spheres.

The basic strategy is to start with the classical expression for
the energy (where we have assumed that $\epsilon$ and ${\mu}$ are 
frequency independent constants)

\begin{eqnarray}
E &=& {1\over2} \int_{Geometry} d^3\vec x
    \left[ \epsilon \vec E^2 + {1\over\mu} \vec { B}^2 \right] .
\end{eqnarray}

\noindent
We now promote the electric and magnetic fields to be operator
quantities, and calculate the vacuum expectation value

\begin{eqnarray}
E = {1\over2} \int_{Geometry} d^3\vec x
    \big[ &\epsilon& \langle  \vec E(0,\vec x) \cdot \vec E(0,\vec x) \rangle
\nonumber\\
  +       &{1\over\mu}& \langle \vec { B}(0,\vec x) \cdot \vec { B}(0,\vec x)
                \rangle \big].
\end{eqnarray}

\noindent
The geometry is incorporated in the calculation both via the limits
of integration and the boundary conditions satisfied by the fields.
Since the above two-point functions are of course divergent, they
must be rendered finite by some regularization prescription.  Milton
{\em et al.}\ use time-splitting (point-splitting in the time
direction), defining the quantity $E(\tau)$ by:

\begin{eqnarray}
\label{E-time-split-1}
E(\tau) = {1\over2} \int_{Geometry} d^3\vec x 
   \big[  
   &\epsilon&   \langle \vec E(\tau,\vec x) \cdot \vec E(0,\vec x) \rangle
\nonumber\\
  +&{1\over\mu}& \langle \vec B(\tau,\vec x) \cdot \vec B(0,\vec x) \rangle 
  \big].
\end{eqnarray}

It should be pointed out that while time-splitting is a very powerful
and technically useful ultra-violet regulator it has the decided
disadvantage of obscuring the underlying physical basis 
of the cut-off in dielectric media.

The technical aspects of the analysis carried out by Milton {\em
et al.}\ reduce then to calculating these two-point correlation
functions (Green functions) by explicitly solving for the TE and
TM modes appropriate for a spherical ball with dielectric boundary
conditions. These Green functions can be written as a sum over
suitable combinations of Ricatti--Bessel functions and vector
spherical harmonics.

To avoid unnecessary notational complications, we schematically
rewrite the above as

\begin{equation} 
E(\tau) = {1\over2} \int d^3\vec x G_{[\epsilon,\mu]}(\tau,\vec x;0,\vec x), 
\end{equation} 

\noindent
where $G_{[\epsilon,\mu]}(t,\vec x;t',\vec x')$ is simply shorthand
for the linear combination of Green functions appearing above.

These Green functions should be calculated for {\em three} different
geometries\footnote{Notice that in Milton {\it et al.} the dielectric
properties of these media are taken to be frequency independent,
with the cut-off being put in via time-splitting.}: \\
{\bf Case I:} A dielectric ball of dielectric constant $\epsilon_1$,
permeability $\mu_1$, and radius $R$ embedded in a infinite dielectric
of {\em different} dielectric constant $\epsilon_2$ and permeability
$\mu_2$. (In applications to sonoluminescence, think of this as an air
bubble of radius $R$ in water.)  \\
{\bf Case II:} A completely homogeneous space completely filled with
dielectric $(\epsilon_2,\mu_2)$. (In applications to sonoluminescence,
think of this as pure water.)  \\
{\bf Case III:} A completely homogeneous space completely filled with
dielectric $(\epsilon_1,\mu_1)$. (In applications to sonoluminescence,
think of this as pure air.)\\


\begin{figure}[htb]
\vbox{\hfil\epsfbox{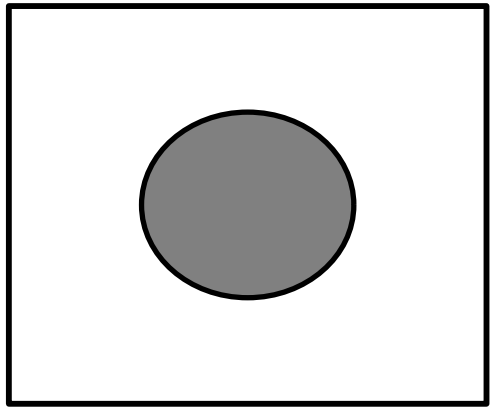}\hfil}
\caption
{\label{F-ball}
Case I: Dielectric ball (foreground) in dielectric background.}
\end{figure}


\begin{figure}[htb]
\vbox{\hfil\epsfbox{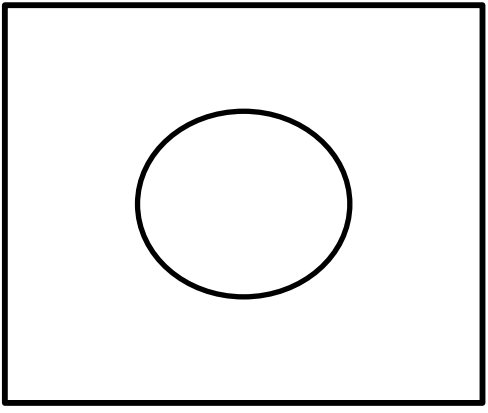}\hfil}
\caption
{\label{F-background}
Case II: Homogeneous dielectric background.}
\end{figure}


\begin{figure}[htb]
\vbox{\hfil\epsfbox{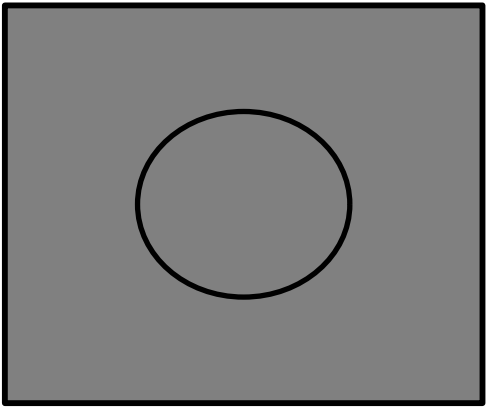}\hfil}
\caption
{\label{F-foreground}
Case III: Homogeneous dielectric foreground.}
\end{figure}

We are in complete agreement with the extant calculations and
results for these three individual Green functions---where we
do not agree, as will be shown, is {\em in the way that these three
Green functions are inserted into the computation for the Casimir
energy.}

Milton {\em et al.}\ calculate an ``energy difference'', which we
will call $E_{\small surface}$, and which they {\em define} as

\begin{equation}
E_{\small surface} = 
{1\over2} \left\{ \int_{\small all~space} G_I - 
                  \int_{r>R} G_{II}  - 
		  \int_{r<R} G_{III} 
\right\}.
\end{equation}

The computation of this quantity is mathematically correct---unfortunately
this is simply not the physically correct quantity to be computing,
and this  definition is equivalent to explicitly excluding by hand
the dominant volume contribution.

The  appropriate physical quantity to compute is~\cite{Schwinger0}

\begin{equation}
E_{\small Casimir} = 
{1\over2} \left\{ \int_{\small all~space} G_I - 
                  \int_{\small all~space} G_{II} \right\}.
\end{equation}

Observe that this quantity is simply the difference in energy
between two  situations: (Case I) having the dielectric ball present
and (Case II) replacing the dielectric ball by the surrounding
medium. {\em This is exactly the quantity that Schwinger calculates
in reference~\cite{Schwinger0}.} Within the context of sonoluminescence
this is the Casimir energy released in evolving from bubble to
no-bubble. In contrast, the definition of $E_{\small surface}$
above is not a physical difference between any two real physical
situations.  Rather it is a doubly subtracted quantity which does
not respect the boundary conditions for the problem it tries to
solve.  In particular, $E_{\small surface}$ is not physically the
same quantity as that calculated by Schwinger.

If we look at the difference between the appropriate definition
(Schwinger's) and that of Milton {\em et al.}\ we see that

\begin{eqnarray}
\Delta E
&=& E_{\small Casimir} - E_{\small surface}
\nonumber\\
&=&
 {1\over2} \int_{r<R} \left\{ G_{III} - G_{II} \right\}.
\label{E-Delta-E}
\end{eqnarray}

This difference is now easily seen to be the missing volume term:
Remember that $G_{III}$ and $G_{II}$ are Green functions corresponding
to spaces completely filled with homogeneous dielectrics---therefore
these Green functions are translation invariant. (When we express
these Green functions in terms of spherical polar coordinates it
is not obvious that they are translation invariant, but because
the physical dielectric in Cases II and III is translation invariant,
the Green functions must also be translation invariant.) This
observation permits us to pull the Green functions outside the
integral, so that

\begin{equation}
\Delta E = {1\over 2} V \left\{ G_{III}(\tau, \vec 0;0, \vec 0) -  
G_{II}(\tau, \vec 0;0, \vec 0) \right\}.
\end{equation}

This explicitly shows that the term omitted in the analyses of Milton
{\em et al.}\ is a volume term. As we show below, it is in fact exactly
the term required to bring that calculation into conformity with
Schwinger's result.

We shall show this by computing the difference term in two separate
ways: (1) using translation invariance, dimensional analysis,
conformal symmetry, and elementary quantum field theory it is
possible to calculate this difference term from first principles
for arbitrary geometries, and (2) particularizing to the case of
a spherical dielectric ball, and working in spherical polar
coordinates, we shall use Milton {\em et al.}'s own formulae for
these Green functions and give an explicit expression for this
difference as a sum over integrals involving Ricatti--Bessel
functions. These integrals and sums will be evaluated explicitly
in closed form, and we shall explicitly see how the correct volume,
cut-off, and dielectric dependence emerge from the calculation.

\bigskip

\section{An elementary quantum field theory calculation}
\label{green}

To evaluate the time-split energy density we apply some standard
results from flat-space Minkowski quantum field theory. Note that
in the Feynman gauge, the two-point function for the electromagnetic
vector potential is given by

\begin{eqnarray}
\langle T( A_\mu(x) A_\nu(y) ) \rangle &=&
{\hbar\over(2\pi)^2} {\eta_{\mu\nu}\over(x-y)^2}
\nonumber\\
&=&
{\hbar\over(2\pi)^2} 
{\eta_{\mu\nu}\over c^2[t(\vec x)-t(\vec y)]^2 - (\vec x-\vec y)^2},
\end{eqnarray}

\noindent and for the Minkowski vacuum

\begin{equation}
\langle \vec E(\tau,\vec x) \cdot \vec E(0,\vec x) \rangle =
\langle \vec B(\tau,\vec x) \cdot \vec B(0,\vec x) \rangle =
{\hbar\over(2\pi)^2} {12\over(c \tau)^4}.
\end{equation}

The (regulated) vacuum energy is  simply [{\em cf.} eq.
(\ref{E-time-split-1})]

\begin{equation}
E(\tau)_{\small vacuum} =
{\hbar\over(2\pi)^2} \; V \; {12\over(c \; \tau)^4}.
\end{equation}

\noindent
However, the effect of adding a frequency independent dielectric
constant, can be simply mimicked by   re-scaling the physical metric
$\eta_{\mu\nu}$, thereby introducing what we shall call the ``optical
metric''

\begin{equation}
g_{\mu\nu}^{\small optical} = {\rm diag}[+1/n^2,-1,-1,-1];
\end{equation}

\begin{equation}
g^{\mu\nu}_{\small optical} = {\rm diag}[+n^2,-1,-1,-1].
\end{equation}

\noindent
Note that on-shell photons in the dielectric will travel along null
curves of this optical metric. The introduction of this optical
metric is a nice technical trick for dealing with homogeneous and
non-dispersive dielectrics by means of viewing them as special
deformations of Minkowski space.

In terms of this optical metric we can still write

\begin{equation}
E(\tau)^{\small optical}_{\small dielectric} =
{\hbar\over(2\pi)^2} \; 
V_{\small optical} \;  
{12\over(c \; \tau_{\small optical})^4}.
\end{equation}

\noindent
This energy is to be interpreted as that which would be measured
by a hypothetical observer for whom the optical metric would be
physical.  This energy must be translated back to physical quantities
by means of  the following equivalences:

\begin{eqnarray}
E_{\small physical} &=& E_{\small optical}/n; \\
\tau_{\small physical} &=& \tau_{\small optical} \; n; \\
V_{\small physical} &=& V_{\small optical}.
\end{eqnarray}

\noindent
The effect of these translations is that the time-split energy
density (time-split in physical time) becomes

\begin{equation}
E(\tau_{\small physical})^{\small physical}_{\small dielectric} =
{\hbar\over(2\pi)^2} V_{\small physical} \; n^3 
{12\over(c \;  \tau_{\small physical})^4}.
\end{equation}

\noindent
But, on the other hand, time-splitting in optical time yields

\begin{equation}
E(\tau_{\small optical})^{\small physical}_{\small dielectric} =
{\hbar\over(2\pi)^2} V_{\small physical} \; {1\over n} \; 
{12\over(c \;  \tau_{\small optical})^4}.
\end{equation}

This now permits us to evaluate {\em exactly} the missing piece
that Milton {\em et al.}\ discarded; with the convention of
time-splitting in physical time, that term is

\begin{eqnarray}
&&\Delta E^{\small physical}_{\small fixed~{\tau_{\small physical}}} 
\nonumber\\
&&\qquad\qquad
= E(\tau_{\small physical})^{\small physical}_{\small dielectric~1} -
E(\tau_{\small physical})^{\small physical}_{\small dielectric~2}
\nonumber\\
&&\qquad\qquad
= {\hbar\over(2\pi)^2} V_{\small physical} \; [n_1^3 - n_2^3 ] \;
 {12\over(c \; \tau_{\small physical})^4}.
\label{E-difference-physical-time}
\end{eqnarray}

Since Milton {\em et al.}\ express all their results in spherical
polar coordinates, as sums over an infinite number of Ricatti--Bessel
functions, it is far from obvious that the rather formidable
expressions encountered in those analyses reduce to the simple and
exact result displayed above. In the next section of this paper we
will do exactly this by invoking a long and turgid agony of special
function theory.

Before leaving this section, however, we wish to emphasize that
comparing this result to the original Schwinger calculation requires
one additional modification: Schwinger regulated his energy
calculation by introducing a cut-off at fixed wave-number
(three-momentum). This is equivalent to time-splitting at fixed
optical time. To see this, recall that frequency and wave-number
cut-offs are in a non-dispersive dielectric related by

\begin{equation}
n \omega_0 = c K,
\end{equation}

\noindent
in which case

\begin{equation}
\tau_{\small physical} \sim 1/\omega_0,
\end{equation}

\noindent
while on the other hand

\begin{equation}
\tau_{\small optical} = 
{\tau_{\small physical}\over n} \sim {1\over n \omega_0}  =
{1\over c K}.
\end{equation}

\noindent
Thus, if we regularize by time-splitting, and perform the subtraction
at fixed optical time, we have

\begin{eqnarray}
&&\Delta E^{\small physical}_{\small fixed~{\tau_{\small optical}}} 
\nonumber\\
&&\qquad\qquad
= E(\tau_{\small optical})^{\small physical}_{\small dielectric~1} -
E(\tau_{\small optical})^{\small physical}_{\small dielectric~2}
\nonumber\\
&&\qquad\qquad
= {\hbar\over(2\pi)^2} V_{\small physical} \left[{1\over n_1} - {1\over n_2}  
\right]
{12\over(c \; \tau_{\small optical})^4}.
\label{E-difference-optical-time}
\end{eqnarray}

This is exactly Schwinger's result as enunciated in~\cite{Schwinger0}.
(More precisely, it is as close as one is ever going to get
considering that a sharp cut-off is just not the same as time-splitting:
There is no reason to expect the dimensionless coefficients to be
the same for the two calculations.  We will comment, near the end
of the next section, how we can switch from a time splitting cut-off
to a sharp cut-off in wave number and obtain exactly Schwinger's
coefficient.)

In summary, what we have shown achieved so far is to show the
following:

\noindent
(1) Schwinger's calculation~\cite{Schwinger0} showing that there
is a bulk Casimir energy in a generic dielectric is correct.

\noindent
(2) The analyses of Milton {\em et al.}~\cite{Milton80,Milton95,Milton96}
omit the volume contribution to the Casimir energy.

\noindent
(3) The volume contribution that was omitted in those
analyses can be calculated exactly, without resorting to special
function theory or asymptotic analysis, and the omitted term exactly
reproduces the volume contribution as evaluated by Schwinger.

\noindent
(4) The volume term in the calculation was missed because of
misidentification of a Green function, due to the incorrect
application of boundary conditions.

\noindent
(5) This argument is not limited to systems of spherical symmetry.
{From} the preceding it is clear that the argument continues to hold
for arbitrarily shaped dielectrics.

In the next section we will evaluate $\Delta E$ given in eq.
(\ref{E-Delta-E}) in yet a different way:  by working directly from
Milton's expression.

\section{The energy difference term in spherical coordinates}
\label{spherical}

{From} the preceding discussion we have isolated the term in the
Casimir energy that Milton {\em et al.}~\cite{Milton80,Milton95,Milton96}
omitted. It is precisely the difference between the Green functions
(of Case III and Case II) integrated over the interior of the bubble
(See eq. (\ref{E-Delta-E})).  We can thus immediately jump into
the middle of the technical computation and directly evaluate this
difference term. This calculation has the advantage of quickly
getting to the heart of the matter.

{From} the two recent papers~\cite{Milton95}, eq.  (41)\footnote{Note
that this is eq. (42) in the hep-th version.}, or from~\cite{Milton96},
eq. (4.2b), we can write

\begin{eqnarray}
\label{E-difference-time-split-definition}
\Delta E &=& E_{\small Casimir} - E_{\small surface}
\nonumber\\
&=& \Re\bigg\{ {-i\over2} \int_{-\infty}^{+\infty}  
{d\omega\over 2\pi}
e^{-i\omega\tau} \int_0^R r^2 dr 
\nonumber\\
&&\qquad \times \left[ \tilde X_{III}(k_{III},r) 
                    -  \tilde X_{II}(k_{II},r) \right]
\bigg\}.
\end{eqnarray}

\noindent
Here we use the notation $k = |\omega| n$, with $n$ the appropriate
refractive index, and define the quantity $\tilde X_{II}(k_{II},r)$
by

\begin{eqnarray}
\tilde X_{II}(k_{II},r) &\equiv&
\sum_{\ell=1}^\infty (2\ell+1)
\Bigg\{ 2 \; k_{II}^2 \; F^{II}_\ell(k_{II};r,r) +  
\nonumber\\
&&+ {1\over r^2} {\partial\over\partial r} \left[ r  
{\partial\over\partial r'} r'
\left[ F^{II}_\ell(k_{II};r,r') \right]\Big|_{r'=r}
\right]
\Bigg\}
\nonumber\\
&&
+ \left[ (F_\ell^{II}) \to (G_\ell^{II}) \right].
\end{eqnarray}

\noindent
The $F_\ell$ and $G_\ell$ Green functions are identical to those
in~\cite{Milton80,Milton95,Milton96}.  Notice that when making the
substitution $(II)\to(III)$ we should also change the refractive
index that implicitly appears in the factor $k$.

In writing these equations we have used the fact that the energy
difference is known to be a {\em real} quantity, so we can immediately
discard the imaginary part of the above expression without bothering
to explicitly evaluate it. There is no subtle physics involved
here---this is just the standard procedure of using complex
exponentials to represent a harmonic time dependence and then taking
the real part at the end of the calculation. In particular there
is no physics hiding in the imaginary part of the above expression.
Not only is there no need to calculate the imaginary part, but it
is physically meaningless to do so.

If we were integrating over all space, then ---
following~\cite{Milton80,Milton95,Milton96} --- the partial derivative
terms in ${\tilde X}(k,r)$ could be safely dropped. Because we are
only integrating over a finite region we must explicitly keep these
derivative terms.

Now because $k= |\omega| n$, while $F_\ell$ and $G_\ell$ are functions
of $k$, they are therefore functions only of the {\em absolute} value
of $\omega$.  This permits us to write

\begin{eqnarray}
\Delta E 
&=& \Re\bigg\{ {-i\over2} \int_{0}^{+\infty} 
{d\omega\over2\pi}
\left[ e^{-i\omega\tau} + e^{+i\omega\tau} \right] 
\nonumber\\
&&\qquad\times \int_0^R r^2 dr  
\left[ \tilde X_{III}(k_{III}r) - \tilde X_{II}(k_{II}r) \right] \bigg\}.
\nonumber\\
&=& \Re\bigg\{ {-i} \int_{0}^{+\infty} {d\omega\over 2\pi} 
\cos(\omega\tau) \;  
\nonumber\\
&&\qquad\times \int_0^R r^2 dr \; 
\left[ \tilde X_{III}(k_{III}r) - \tilde X_{II}(k_{II}r) \right]  \bigg\},
\end{eqnarray}

\noindent
and  finally

\begin{eqnarray}
\label{E-difference-time-split-intermediate}
\Delta E &=& \int_{0}^{+\infty} {d\omega\over 2\pi}
\cos(\omega\tau) \; \int_0^R r^2 dr \; 
\nonumber\\
&&
\times
\Im\ \left[ \tilde X_{III}(k_{III},r) - \tilde X_{II}(k_{II},r) \right].
\nonumber\\
\end{eqnarray}

The Green functions  for Cases II and III are very simple (since
we are dealing with homogeneous spaces) and the imaginary parts
can easily be read off from~\cite{Milton80} eq. (29), \cite{Milton95}
eq. (35), or~\cite{Milton96} eq. (3.7)\footnote{Note: The two lines
given in those equations are individually the Green functions for
Cases II and III (defined over the whole space). The combination
given in~\cite{Milton80,Milton95,Milton96} is {\em not\/} a Green
function of {\em any} differential operator. See the discussion
surrounding eq. $(\ref{E-Delta-E})$ earlier in this paper.}

\begin{eqnarray}
\Im\{F_\ell\} = \Im\{G_\ell\} &=& \Im\{ i k \; j_\ell(k,r) \; h_\ell(kr) \}
\nonumber\\
&=& k \; j_\ell^2(kr) 
\nonumber\\
&=& k \; {s_\ell^2(kr)  \over  (kr)^2}.
\end{eqnarray}

\noindent
Here we have introduced the Ricatti--Bessel function $s_\ell(x)$.
(We refer the reader to the Appendix for details.) Inspection  of
the derivative pieces yields

\begin{eqnarray}
&&
\Im\left\{{\partial\over\partial r} \left[ r {\partial\over\partial  r'} r'
\left[ F_\ell(k;r,r') \right]\Big|_{r'=r} \right] \right\}
\nonumber\\
&&\qquad=
\Im\left\{ {\partial\over\partial r} \left[r {\partial\over\partial  r'} r'
\left[ G_\ell(k;r,r') \right]\Big|_{r'=r} \right]\right\}
\nonumber\\
&&\qquad=
\Im\big\{ i k \; \big[ (x h_\ell(x))'|_{kr} \; (x j_\ell(x))'|_{kr}
\nonumber\\
&&\qquad\qquad +  (x h_\ell(x))|_{kr} \; (x j_\ell(x))''|_{kr} \big] \big\}
\nonumber\\
&&\qquad=
k \; \Re\{  e_\ell'(x)|_{kr} \; s_\ell'(x)|_{kr} +  e_\ell(x)|_{kr}  
\; s_\ell''(x)|_{kr} \}
\nonumber\\
&&\qquad=
k \; \{  s_\ell'(x)|_{kr} \; s_\ell'(x)|_{kr} +  s_\ell(x)|_{kr} \;  
s_\ell''(x)|_{kr} \},
\end{eqnarray}
with  $e_\ell(x)$ another Ricatti-Bessel function.

To derive these results we can either use brute force starting from
eq. (35) of~\cite{Milton95}, or alternatively we could inspect
equations (19--22) of~\cite{Milton80}.  (Note that there was an
overall change in normalization between the 1980 paper and the 1995
and 1996 papers, and take account of the fact that we are here
retaining the total derivative term.)

All together, this implies

\begin{eqnarray}
&&\Im\left[ \tilde X_{III}(k_{III},r) - \tilde X_{II}(k_{II},r)\right] 
\nonumber\\
&&=
2\;  {k_{III}\over r^2} \; \sum_{\ell=1}^\infty (2\ell+1)
\nonumber\\
&&\times 
\left\{ 2 (s_\ell(x))^2 + ([s_\ell(x)]')^2 + s_\ell(x) s_\ell''(x)  
\right\}|_{k_{III}r}
\nonumber\\
&& - [ (III) \to (II) ].
\end{eqnarray}

The sum over Ricatti--Bessel functions is now easily and {\em
exactly} evaluated using the results in the Appendix. We
simply get

\begin{eqnarray}
\Im\{ \tilde X_{III}(k_{III},r) - \tilde X_{II}(k_{II},r) \} 
&=&
2 \; \left[ {k\over r^2}\; 2 (kr)^2\right] \Big|^{(III)}_{(II)}
\nonumber\\
&=& 4 \left[ k_{III}^3 - k_{II}^3 \right]
\nonumber\\
&=& 4 \left[ n_1^3 - n_2^3 \right] |\omega|^3.
\end{eqnarray}

\noindent
Notice that the above expression  is independent of $r$ ---a result
that is by no means obvious from the original definition. (But we
knew from the fact that the underlying Green functions are translation
invariant that this quantity had to be independent of position at
the end of the day!)

Turning again to the energy difference

\begin{equation}
\label{E-difference-time-split-penultimate}
\Delta E = \left[ n_1^3 - n_2^3 \right]
 \int_{0}^{+\infty} {d\omega\over2\pi}
\left[ \cos(\omega\tau) \right] \int_0^R r^2 dr \; 4  \omega^3.
\end{equation}

\noindent
Which is now easily evaluated, using $x= \omega\tau$, as

\begin{equation}
\Delta E =  {2\over3\pi} \left[ n_1^3 - n_2^3 \right]  
\; {R^3\over\tau^4}
\int_{0}^{+\infty}
x^3 \cos(x) dx.
\end{equation}

\noindent
As a penultimate step we use 

\begin{equation}
\int_{0}^{+\infty} x^3 \cos(x) dx = 6,
\end{equation}

\noindent
and $V = (4\pi/3) R^3$, to write

\begin{equation}
\Delta E =  {3\over\pi^2} \left[ n_1^3 - n_2^3 \right]  
{1\over\tau^4} \; V.
\end{equation}

\noindent
Re-inserting factors of $\hbar$ and $c$, which have been suppressed
for clarity, we see

\begin{equation}
\Delta E(\tau_{\small physical}) =  {3\over\pi^2} \; \hbar c \;
\left[ n_1^3 - n_2^3 \right] {1\over( c \; \tau_{\small physical})^4} \; V.
\end{equation}

\noindent
Note that this is exactly the same result ({\em including numerical
coefficients}) as  the  previously calculated from general considerations
using translation invariance and the strategy of introducing the
optical metric. [See eq. (\ref{E-difference-physical-time}).]

Finally, as discussed previously, we note that to compare this
result to Schwinger's we should be time-splitting in optical time,
and so must absorb a few factors of refractive index into the
time-splitting parameter. [See eq. (\ref{E-difference-optical-time}).]

\begin{equation}
\Delta E(\tau_{\small optical}) = {3\over \pi^2}\; \hbar c \;
\left[{1\over n_1} - {1\over n_2} \right] {1\over(c\;  
\tau_{\small optical})^4} V.
\end{equation}

\noindent
This is exactly the time-split version of Schwinger's expression
(\ref{E-difference-physical-time}) for the bulk contribution to
the Casimir energy.

We have thus evaluated the difference term, $\Delta E = E_{\small Casimir}
- E_{\small surface}$, in two independent and completely independent
ways. The two calculations agree down to exact numerical coefficients.
Turning this result around, we have

\begin{equation}
E_{\small Casimir} = \Delta E + E_{\small surface}.
\end{equation}

\noindent
So we see that the term we have just calculated provides  the
bulk volume contribution to the Casimir energy that is required to
bring Milton {\em et al.}'s  calculations into conformity with Schwinger's
arguments~\cite{Schwinger0}, and into conformity with the general
arguments provided in this paper.

To finally compare the {\em absolute normalization} of this result with
Schwinger's, we must explicitly replace the time-splitting by a
wave-number cut-off. Simply go back to eq.
(\ref{E-difference-time-split-intermediate}), introduce a wave-number
cut-off, and let the time-splitting parameter go to zero, to obtain

\begin{eqnarray}
\label{E-difference-wave-number-intermediate}
\Delta E &=& \int_{0}^{+\infty} {d\omega\over 2\pi}\; \int_0^R r^2 dr \; 
\nonumber\\
&&
\times
\Im\{f(k_{III})\tilde X_{III}(k_{III},r) 
   - f(k_{II})\tilde X_{II}(k_{II},r)\}.
\nonumber\\
\end{eqnarray}

\noindent
The function $f(k)$ is any real smooth function with $f(0)=1$ and
$f(+\infty) = 0$. For simplicity we have temporarily used the {\em
same} cut-off function for the two media.  We generalize the argument
below.  The evaluation of the various pieces of the Green functions
proceeds as before, so that eq. now
(\ref{E-difference-time-split-penultimate}) becomes

\begin{equation}
\label{E-difference-wave-number-penultimate}
\Delta E =
\int_{0}^{+\infty} {d\omega\over2\pi}
\int_0^R r^2 dr 4 \{f(k_{III}) k_{III}^3 - f(k_{II}) k_{II}^3\}.
\end{equation}

\noindent
After a change of variables, and an explicit evaluation of the volume
integral, we easily get

\begin{equation}
\label{E-difference-wave-number-ultimate}
\Delta E =
\left[{1\over n_1} - {1\over n_2}\right] V 
\int{d^3 \vec k\over(2\pi)^3} f(k) k.
\end{equation}

\noindent
Approximating air by vacuum, (i.e. setting $n_1=1$), and inserting
a sharp cut-off at wave-number $K$, (by setting $f(k) =
\Theta(K-k)$), this is {\em exactly} Schwinger's bulk volume term
for a cavity in a dielectric [eq. (\ref{E-cavity-in-dielectric})]
down to the last numerical prefactor.

Now the use of a single cut-off function, while mathematically more
transparent, is physically dubious.  If we introduce separate
cut-off functions for the two media the modifications are
straightforward. First

\begin{eqnarray}
\label{E-difference-wave-number-intermediate-2}
\Delta E &=& \int_{0}^{+\infty} {d\omega\over 2\pi}\; \int_0^R r^2 dr \; 
\nonumber\\
&&
\times
\Im\{f_{III}(k_{III})\tilde X_{III}(k_{III},r) 
   - f_{II}(k_{II})\tilde X_{II}(k_{II},r)\}.
\nonumber\\
\end{eqnarray}

\noindent
The functions $f_a(k)$ are now any two real smooth functions with
$f_a(0)=1$ and $f_a(+\infty) = 0$. We now have

\begin{equation}
\label{E-difference-wave-number-penultimate-2}
\Delta E =
 \int_{0}^{+\infty} {d\omega\over2\pi}
\int_0^R r^2 dr 4 \{f_{III}(k_{III}) k_{III}^3 - f_{II}(k_{II}) k_{II}^3\}.
\end{equation}

\noindent
After a change of variables, setting  $f_{III} = f_1$ and $f_{II}
= f_2$, and explicit evaluation of the volume integral, we easily
get

\begin{equation}
\label{E-difference-wave-number-ultimate-2}
\Delta E =
V \int{d^3 \vec k\over(2\pi)^3} 
\left[ {f_1(k)\over n_1} -{f_2(k)\over n_2}\right] k.
\end{equation}
\noindent
Inserting sharp cut-offs in wave-number, \'a la Schwinger, 
is accomplished 
by setting

\begin{equation}
f_a(k) = \left[\Theta(K_a-k) + n_a \Theta(k-K_a)\right] \Theta(K_\infty-k).
\end{equation}

\noindent
These cut-off functions may look a little mysterious, depending as
they do on {\em three} cut-off scales $K_1$, $K_2$, and $K_\infty$.
The two cut-offs $K_1$ and $K_2$ are physical---they describe the
wave-numbers at which the two dispersion relations effectively
approach the vacuum dispersion relation. The third cut-off $K_\infty$
is purely a mathematical convenience to make the integrals converge.
$K_\infty$ should be taken to be much larger than either $K_1$ or
$K_2$. With this notation

\begin{eqnarray}
{ c k f_a(k)\over n_a} &=& 
\left[{c k\over n_a} \Theta(K_a-k) + c k  \Theta(k-K_a)\right] 
\Theta(K_\infty-k)
\nonumber\\
&=&
\omega_a(k) \Theta(K_\infty-k).
\end{eqnarray}

\noindent
That is to say, these particular cut-off functions have been chosen
to mimic the model dispersion relation used in the mode sum discussion
(section \ref{modesum}).

Evaluating the integrals, and reinserting factors of $\hbar$ and
$c$, we can write

\begin{equation}
\Delta E = 
{1\over8\pi^2} V \; \hbar c \; 
\left\{ K_1^4 \left[ {1\over  n_1} - 1 \right] -  
        K_2^4 \left[ {1\over  n_2} - 1 \right] \right\}.
\label{E-two-cut-offs-2}
\end{equation}

\noindent
Note in particular that $K_\infty$ has quietly disappeared from
this answer.  This is {\em exactly} the same as the result obtained
by mode sum arguments in section \ref{modesum} down to the last
numerical prefactor.  [See eq. (\ref{E-two-cut-offs}).]

A slightly more general cut-off is to use the wave-number dependent
refractive index to construct the cut-off according to the scheme

\begin{equation}
f_a(k) = \left[{n_a(k=0)\over n_a(k)}\right] \Theta(K_\infty-k).
\end{equation}

\noindent
In which case, recalling that that $n_{a} = n_{a}(k=0)$

\begin{equation}
{ c k f_a(k)\over n_a} = 
\left[{c k\over n_a(k)} \right] \Theta(K_\infty-k) =
\omega_a(k) \Theta(K_\infty-k).
\end{equation}

\noindent
The Casimir energy difference, including reinserted factors of 
$\hbar$ and $c$, is now

\begin{equation}
\label{E-difference-wave-number-ultimate-3}
\Delta E =
\hbar V \int{d^3 \vec k\over(2\pi)^3} 
\left[ \omega_1(k) -\omega_2(k)\right] \Theta(K_\infty-k).
\end{equation}

\noindent
We can now let $K_\infty$ tend to infinity, and thereby recover
eq. (\ref{E-two-dielectrics}).

In summary, the energy difference term calculated via explicit
summation over Ricatti--Bessel functions is completely in agreement
with the dominant volume term coming from simple mode sum and
density of states arguments.

\section{Full calculation from first principles}
\label{full}

\subsection{Background}

There is also a certain amount of value to going to the trouble of
redoing the calculation from first principles.

Recall that the physical quantity that we are interested in evaluating
for the calculation of the Casimir energy is the total energy. The
total classical energy is computed from the integral of the energy
density over the geometry\footnote{This expression is valid for
frequency independent  $\epsilon$ and $\mu$ only.},

\begin{equation}
E=\int_{Geometry} d^3 \vec{x} \; \frac{1}{2} \;
\left( \epsilon \vec{E}^2  
+ \mu \vec{H}^2 \right).
\end{equation}

\noindent
After quantization this energy is related to the vacuum expectation
value of the squares of the electric and magnetic fields in the
dielectric.

Since the electric and magnetic fields  are related to derivatives
of the electromagnetic potential, the energy density is calculable
in terms of the two--point correlation function of the electromagnetic
field; these correlators are in general divergent, and we need to
perform renormalizations before comparing with the results of
experiments.  This, of course, is nothing but the general recipe
for evaluating quantities in any quantum field theory.

This calculation has been done in detail by Milton {\em et
al.}~\cite{Milton80,Milton95,Milton96} for the geometrical
configuration previously described. They computed the total energy $E$
by integrating the electromagnetic energy density over the geometry
and then rendered the expressions finite using point-splitting
regularization in the time direction. We quote the expressions from
them (see below for the various definitions), as

\begin{eqnarray}
E &=&
\Re\Bigg\{ 
\int d^3 \vec{x} \; \frac{1}{2} \; \left( \epsilon \vec{E}^2 +  
\mu \vec{H}^2 \right) \Bigg\}
\nonumber\\
\label{E-Milton-energy}
&=&\Re\Bigg\{ \frac{1}{2i} \int_{-\infty}^{\infty} 
\frac{d \omega}{2 \pi} e^{-i  \omega (t-t')} 
\sum_{l=1}^{\infty} (2l+1)
\nonumber\\
&&\qquad\times \int_{0}^{\infty} r^2 dr 
\big\{
2 k^2 [ F_\ell(k;r,r) + G_\ell(k;r,r)] 
\nonumber\\
&&\qquad\qquad + [{\rm Derivative\;\; Terms}]
\big\} \Bigg\}.
\end{eqnarray}

\noindent
Here the total derivative can be shown (Ref. 10 of Milton) to
integrate to zero.\footnote{We reiterate that this vanishing of
the total derivative terms depends crucially on the continuity and
differentiability of the Green functions.}  In this equation the
$\omega$--integral is for the regulator in the time--direction,
and the radial functions $F_\ell (r,r')$ and $G_\ell (r,r')$ are
related to the Green functions for the electric and magnetic fields
in the appropriate geometry. For a dielectric sphere of radius $R$,
permeability $\epsilon_1$ and permittivity $\mu_1$, embedded in
another dielectric with $\epsilon_2$ and $\mu_2$, using the results
of Boyer~\cite{Boyer,Boyer2} as adapted by Milton {\em et al.}, it
is possible to readily check that these objects are the radial part
of the Green functions for the wave equation in spherical coordinates,

\begin{eqnarray}
&&\left( {\partial^2 \over \partial r^2}    + \frac{2}{r}      
{\partial \over \partial r}  - \frac{\ell (\ell +1)}{r^2} + k^2  
 \right) F_{\ell}(k;r,r')
\nonumber\\
&&\qquad\qquad=-\frac{1}{r^2} \delta(r-r'),
\label{E-d-e}
\end{eqnarray}

\noindent
and are given explicitly as follows:

\noindent
For $r,r'<R$,

\begin{eqnarray}
\label{E-Milton-F}
F_\ell (k_{\small in};r,r') &=& i k_{\small in} j_\ell (k_{\small in} r_<)
\nonumber\\
&&\times
[ h_\ell (k_{\small in} r_>)- A^\ell_F j_\ell (k_{\small in} r_>)].
\end{eqnarray}

\noindent
For $r,r'>R$,

\begin{eqnarray}
F_\ell (k_{\small out};r,r') &=& i k_{\small out} h_\ell (k_{\small out} r_>)
\nonumber\\
&&\times
[ j_\ell (k_{\small out} r_<)- B^\ell_F h_\ell (k_{\small out} r_<)].
\end{eqnarray}

\noindent
A similar expression holds for $G_\ell$. In these two expressions
the wave numbers are $k_{\small in}=|\omega| \sqrt{\epsilon_1
\mu_1}$ and $k_{\small out} =|\omega| \sqrt{\epsilon_2 \mu_2}$.
The function $j_\ell (x)$ is the spherical Bessel function of order
$\ell$ and $h_\ell (x) \equiv h_\ell^{(1)} (x)$ is the spherical
Hankel function of the first kind. The quantities $A^\ell_F$,
$A^\ell_G$, $B^\ell_F$ and $B^\ell_G$, are given
in~\cite{Milton80,Milton95,Milton96} and their explicit forms are
obtained by requiring that the Green functions be solutions to the
appropriate equations satisfying the correct boundary conditions.

Milton {\em et al.}\ in their papers give the expression for the
energy that we quote above, eq. (\ref{E-Milton-energy}), after they
have integrated over all of space before subtracting the energies
for each material configuration. This is dangerous since there is
then an additional infrared divergence arising from the volume
integral which is not regulated by the time splitting--procedure,
which only regulates UV divergences. To avoid these unnecessary
pitfalls we will write down the energy density, perform the
subtractions at the level of the densities (local subtractions)
and then integrate.

Using the spherical Bessel function form for the electric and
magnetic fields satisfying the Maxwell equations, the energy density,
$T^{tt}$, is seen to be given by

\begin{equation}
T^{tt}(r) = 
\Re\bigg\{ {-i\over2} \int_{-\infty}^{+\infty} {d\omega\over 2\pi} 
e^{-i\omega\tau} \; X(k,r) \bigg\},
\end{equation}

\noindent
where we have used the notation $k = |\omega| n$ with $n$ the
refractive index appropriate for the medium, and defined the quantity
$X(k,r)$ via\footnote{The quantity $X(k,r)$ defined here is (apart
from a physically irrelevant $\delta(0)$ term, see below) equal to
the quantity $\tilde X(k,r)$ defined in the previous section---these
quantities differ only by the way that the derivative terms have
been manipulated.}

\begin{eqnarray}
X(k,r) &\equiv&
\sum_{\ell=1}^\infty (2\ell+1)
\Bigg\{ \left[ k^2 + {\ell(\ell+1)\over r^2} \right]
\left[ F_\ell(k;r,r)\right]
\nonumber\\
&&+ {1\over r^2} {\partial\over\partial r_1} r_1  
{\partial\over\partial r_2} r_2
 \left[ F_\ell(k;r_1,r_2)  
\right]\Big|_{r_1=r_2=r}
\Bigg\},
\nonumber\\
&& + [ (F_\ell) \to (G_\ell) ].
\end{eqnarray}

\noindent
This is of course equivalent to eq. (\ref{E-Milton-energy}) above.
Here $k_{I} \equiv |\omega| n_{I}$ represents the wave-number for
light of frequency $\omega$ in a medium with index of refraction
$n_{I}$. Similar results, with appropriate substitutions for the
momenta and Green functions, hold for the situations when we make
reference to Cases II and III.  Note that when making the substitutions
$(I)\to(II)$ or $(I)\to(III)$ we must also change the refractive
index that implicitly appears in the factor $k$. In addition, it
should be borne in mind that $k_I$ is a function of position: $k_I
= n_1 |\omega|= k_{III}$ inside the dielectric sphere, whereas $k_I
= n_2 |\omega| = k_{II}$ outside the dielectric sphere.

The Casimir energy is now obtained by taking the {\em difference}
in energy densities and integrating over all of space while paying
attention to the appropriate index of refraction for each region
of space. For example, the Casimir energy difference between Cases
I and II is given by

\begin{eqnarray}
\label{E-Casimir-energy}
E_{\small Casimir} &=& \Re\bigg\{ {-i\over2} \int_0^\infty r^2 dr \;
\int_{-\infty}^{+\infty} {d\omega\over 2\pi}
\nonumber\\
&&
\qquad\times e^{-i\omega\tau} X(k_{I},k_{II},r) \bigg\},
\end{eqnarray}

\noindent
where we have defined the quantity $X(k_{I},k_{II},r)$ by

\begin{eqnarray}
X(k_{I},k_{II},r) &\equiv& X_I(k_I,r) - X_{II}(k_{II},r)
\nonumber\\
&=&
\sum_{\ell=1}^\infty (2\ell+1)
\Bigg\{ \left[ k_{I}^2 + {\ell(\ell+1)\over r^2} \right]
\nonumber\\
&& \times
\left[ F^{I}_\ell(k_{I};r,r) + G^{I}_\ell(k_{I};r,r) \right]
\nonumber\\
&&+ {1\over r^2} \; {\partial\over\partial r_1} r_1  
{\partial\over\partial r_2} r_2
\big[ F^{I}_\ell(k_{I};r_1,r_2) 
\nonumber\\
&& \qquad + G^{I}_\ell(k_{I};r_1,r_2)  \big]\Big|_{r_1=r_2=r}
\Bigg\}
\nonumber\\
&&
- \left[ (I) \to (II) \right].
\end{eqnarray}

\noindent
To turn this into eq. (41) of reference~\cite{Milton95}  
requires a few technicalities which were not made explicit by Milton
but we will show here. (Note that the above is his  
eq. (42) in the hep-th version of the paper.)

Observe that
\begin{eqnarray}
&&{\partial\over\partial r} \left[ r {\partial\over\partial r'} r'
 \left[ F^{I}_\ell(k_{I};r,r') + G^{I}_\ell(k_{I};r,r')  
\right]\Big|_{r'=r} \right]
\nonumber\\
&&=
 {\partial\over\partial r_1} r_1 {\partial\over\partial r_2} r_2
 \left[ F^{I}_\ell(k_{I};r_1,r_2) + G^{I}_\ell(k_{I};r_1,r_2)  
\right]\Big|_{r_1=r_2=r}
\nonumber\\
&&\qquad+
  r {\partial^2\over\partial r'^2} r'
 \left[ F^{I}_\ell(k_{I};r,r') + G^{I}_\ell(k_{I};r,r')  
\right]\Big|_{r'=r}
\nonumber\\
&&=
 {\partial\over\partial r_1} r_1 {\partial\over\partial r_2} r_2
 \left[ F^{I}_\ell(k_{I};r_1,r_2) + G^{I}_\ell(k_{I};r_1,r_2)  
\right]\Big|_{r_1=r_2=r}
\nonumber\\
&&\qquad-
 \left[ k_{I}^2 - {\ell(\ell+1)\over r^2} \right] r^2
 \left[ F^{I}_\ell(k_{I};r,r) + G^{I}_\ell(k_{I};r,r) \right]
\nonumber\\
&&\qquad- 2 \delta(0),
\end{eqnarray}

\noindent
where in the last line we have used the differential eq. (\ref{E-d-e})
satisfied by the Green functions to  replace the double derivative
at the same point by an explicit polynomial.

The term proportional to $\delta(0)$ arises from the source term
on the RHS of the differential equation (\ref{E-d-e}) defining the
Green functions. Fortunately this term is completely independent
of the dielectric properties and in fact is independent of all
properties of the medium.  Thus when calculating energy {\em
differences} this term cancels identically.  (Earlier calculations
often quietly discard this term without even mentioning its
existence.)

Inserting this into the general formula for the Casimir energy, eq.
(\ref{E-Casimir-energy}), yields eq. (41) of reference~\cite{Milton95}.
Explicitly, it allows us to re-write $X(k_I,k_{II},r)$ as

\begin{eqnarray}
X(k_{I},k_{II},r) &\equiv&
\sum_{\ell=1}^\infty (2\ell+1)
\Bigg\{ 2 k_{I}^2
\Big[ F^{I}_\ell(k_{I};r,r) 
\nonumber\\
&&\qquad\qquad 
+ G^{I}_\ell(k_{I};r,r) \Big]
\nonumber\\
&&+ {1\over r^2} \; {\partial\over\partial r} \Bigg[ r  
{\partial\over\partial r'} r'
\Big[ F^{I}_\ell(k_{I};r,r') 
\nonumber\\
&&\qquad\qquad
+ G^{I}_\ell(k_{I};r,r') \Big]
\Big|_{r'=r} \Bigg]
\Bigg\}
\nonumber\\
&&
- \left[ (I) \to (II) \right],
\end{eqnarray}

\noindent
which is now straightforward to evaluate.

We know that $k$ is only dependent on the {\it absolute} value of
$\omega$, which permits us to write

\begin{eqnarray}
E_{\small Casimir} &=& 
\Re \Bigg\{ {-i\over2}  \int_0^\infty r^2 dr \int_{0}^{+\infty}  
{d\omega\over 2\pi}
\left[ e^{-i\omega\tau} + e^{+i\omega\tau} \right]  
\nonumber\\
&& 
\qquad \times X(k_{I},k_{II},r) \Bigg\}.
\end{eqnarray}

\noindent
Wherefore

\begin{eqnarray}
E_{\small Casimir} 
&=& 
\Re \Bigg\{{-i}  \int_0^\infty r^2 dr \; \int_{0}^{+\infty}  
{d\omega\over 2\pi}
\cos(\omega\tau) \; 
\nonumber\\
&&
\qquad  \times X(k_{I},k_{II},r) \Bigg\}.
\end{eqnarray}

\noindent
We are thus led to

\begin{eqnarray}
\label{E-Casimir-time-split-intermediate}
E_{\small Casimir} 
&=&  
\int_0^\infty r^2 dr \; \int_{0}^{+\infty}  {d\omega\over 2\pi}
\nonumber\\
&&
\qquad\times
\cos(\omega\tau) \; \Im\{X(k_{I},k_{II},r)\}.
\end{eqnarray}

\noindent
With {\em due caution}, the relevant pieces of the Green functions
can be read off from~\cite{Milton80,Milton95,Milton96}.

\subsection{Calculation of the various Green functions.}

The general form of the Green functions were given above, but one
must require that the boundary conditions appropriate to the geometry
and the physics be correctly incorporated into them. This means
that they must satisfy appropriate continuity conditions derived
from Maxwell's equations. In terms of the fields,

\begin{equation}
\vec E_\perp, \quad \epsilon \vec E_r, \quad
{1\over\mu}\vec B_\perp, \quad {\rm and} \quad \vec B_r,
\end{equation}

\noindent
must be continuous. In terms of the radial Green functions, $F_{\ell}$
and $G_{\ell}$,

\begin{equation}
\mu F_{\ell}, \quad G_{\ell}, \quad
{\partial \over \partial r}{r F_{\ell}}, \quad {\rm and} \quad
\frac{1}{\epsilon}{\partial \over \partial r}{r G_{\ell}},
\end{equation}

\noindent
must be continuous.

We now deal with each of the above three cases.

\noindent
{\bf Case I:} \\
Consider a spherically symmetric bubble of radius $R$ and index of
refraction $n_1$ embedded in an otherwise homogeneous material of
index of refraction $n_2$.  (See figure \ref{F-ball}).  The Green
functions satisfying the correct boundary conditions are:

\noindent
For $r_1, r_2 < R$,

\begin{eqnarray}
F^I_\ell, G^I_\ell(r_1,r_2) &=&
i k_{III} \; j_\ell(k_{III} r_<) \;
\nonumber\\
&& \times
\left[h_\ell(k_{III}r_>) - A^\ell_{F,G} \; j_\ell(k_{III} r_>) \right],
\end{eqnarray}

\noindent
For $r_1, r_2 >R$,

\begin{eqnarray}
F^I_\ell, G^I_\ell(r_1,r_2) &=&
i k_{II} \; h_\ell(k_{II} r_>) \;
\nonumber\\
&& \times
\left[j_\ell(k_{II}r_<) - B^\ell_{F,G} \; h_\ell(k_{II} r_<) \right].
\end{eqnarray}

\noindent
(See equations (12a) and (12b) of~\cite{Milton80}, and eq. (16)
of~\cite{Milton95}.) The quantities $A^\ell_{F,G}$ and $B^\ell_{F,G}$
are the ones given by Milton {\em et
al.}~\cite{Milton80,Milton95,Milton96}.

\noindent
{\bf Case II:} \\
For the configuration that we have called Case II  (see
figure \ref{F-background}), the Green function is:  

\noindent
For all $r_1,r_2$

\begin{equation}
F^{II}_\ell, G^{II}_\ell(r_1,r_2) = 
i k_{II} \; j_\ell(k_{II} r_<) \; h_\ell(k_{II}r_>).
\end{equation}

\noindent

\noindent
{\bf Case III:} \\
Finally for Case III (see figure \ref{F-foreground}) the Green
function is:

\noindent
For all $r_1, r_2$

\begin{equation}
F^{III}_\ell, G^{III}_\ell(r_1,r_2) = 
i k_{III} \; j_\ell(k_{III} r_<) \; h_\ell(k_{III}r_>).
\end{equation}

\noindent
We are now ready to explicitly compute the Casimir energy.

The imaginary parts can now easily be read off. For Case I with
$r<R$, some straightforward algebraic manipulations lead to:

\begin{eqnarray}
\Im\{F^I_\ell\}(k_{III};r,r) &=& \Im\{G^I_\ell\}(k_{III};r,r)
\nonumber\\
&=&
\Re\Big\{ k_{III} \; j_\ell(k_{III}r) \;
\nonumber\\
&&\times 
\left[h_\ell(k_{III}r) - A^\ell_{F,G} \; j_\ell(k_{III} r)  
\right] \Big\}
\nonumber\\
&=& (1-\Re\{A^\ell_{F,G}\}) \; k_{III}  \; j_\ell^2(k_{III}r) \;  
\nonumber\\
&=& (1-\Re\{A^\ell_{F,G}\}) \; k_{III} \; 
{s_\ell^2(k_{III}r) \; \over  (k_{III}r)^2},
\end{eqnarray}

\noindent
where to simplify the writing we have again introduced the
Ricatti--Bessel functions. (See the Appendix.)

Similarly, for $r>R$ we get
\begin{eqnarray}
\Im\{F^I_\ell\}(k_{II};r,r) &=& \Im\{G^I_\ell\}(k_{II};r,r)
\nonumber\\
&=&
\Re\Big\{ k_{II} \; h_\ell(k_{II}r) \;
\nonumber\\
&&\times   \left[j_\ell(k_{II}r) - B^\ell_{F,G} \; h_\ell(k_{II} r) \right]  
\Big\}
\nonumber\\
&=& k_{II}  \; j_\ell^2(k_{II}r) -
  k_{II} \; \Re\left \{B^\ell_{F,G} \; h_\ell^2(k_{II} r)\right\}
\nonumber\\
&=& k_{II} \; {s_\ell^2(k_{II}r)  \over  (k_{II}r)^2} -
  k_{II} \; {\Re\left \{B^\ell_{F,G} \; e_\ell^2(k_{II} r)\right\} \over  
(k_{II} r)^2 }.
\nonumber\\
\end{eqnarray}

\noindent
On the other hand, for Case II we simply have

\begin{eqnarray}
\Im\{F^{II}_\ell\}(k_{II};r,r) &=& \Im\{G^{II}_\ell(k_{II};r,r)\}
\nonumber\\
&=&
\Re\left \{ k_{II} \; j_\ell(k_{II}r) \; h_\ell(k_{II}r) \right\}
\nonumber\\
&=& k_{II}  \; j_\ell^2(k_{II}r) 
\nonumber\\
&=& k_{II} \; {s_\ell^2(k_{II}r) \over  (k_{II}r)^2}.
\end{eqnarray}

\noindent
Inspection of the derivative pieces yields, for $r<R$:

\begin{eqnarray}
D^I_{\small in}
&\equiv&
\Im\left\{{\partial\over\partial r_1} r_1 {\partial\over\partial r_2} r_2
          \left[ F^I_\ell(r_1,r_2) \right]\Big|_{r_1=r_2=r} \right\}
\nonumber\\
&\equiv&
\Im\left\{{\partial\over\partial r_1} r_1 {\partial\over\partial r_2} r_2
\left[ G^I_\ell(r_1,r_2) \right]\Big|_{r_1=r_2=r} \right\}
\nonumber\\
&=&
\Im\{ i k_{III} \; (x j_\ell(x))'|_{k_{III}r} \;
\nonumber\\
&&\times \left[ (x h_\ell(x))'|_{k_{III}r} - A^\ell_{F,G} \; (x  
j_\ell(x))'|_{k_{III}r} \right]\}
\nonumber\\
&=&
k_{III} \; \Re\{  s_\ell'(x)|_{k_{III}r} \;
\nonumber\\
&& \times 
   \left[ e_\ell'(x)|_{k_{III}r} - 
          A^\ell_{F,G} \; s_\ell'(x)|_{k_{III}r}  
   \right] \}
\nonumber\\
&=&
(1-\Re\{A^\ell_{F,G}\}) \; k_{III} \; \{  [s_\ell'(x)]^2|_{k_{III}r} \}
\end{eqnarray}

\noindent
While for $r>R$ we have:

\begin{eqnarray}
D^I_{\small out}
&\equiv&
\Im\left\{{\partial\over\partial r_1} r_1 {\partial\over\partial r_2} r_2
\left[ F^I_\ell(r_1,r_2) \right]\Big|_{r_1=r_2=r} \right\}
\nonumber\\
&\equiv&
\Im \left\{{\partial\over\partial r_1} r_1 {\partial\over\partial  
r_2} r_2
\left[ G^I_\ell(r_1,r_2) \right]\Big|_{r_1=r_2=r} \right\}
\nonumber\\
&=&
\Im\{ i k_{II} \; (x h_\ell(x))'|_{k_{II}r} \;
\nonumber\\
&&\times 
\left[ (x j_\ell(x))'|_{k_{II}r} - 
       B^\ell_{F,G} \; (x h_\ell(x))'|_{k_{II}r} 
\right]\}
\nonumber\\
&=&
k_{II} \; \Re\{  e_\ell'(x)|_{k_{II}r} \;
\nonumber\\
&&\times
\left[ s_\ell'(x)|_{k_{II}r} - B^\ell_{F,G} \; e_\ell'(x)|_{k_{II}r}  
\right] \}
\nonumber\\
&=&
k_{II} \; \{  s_\ell'(x)|_{k_{II}r} \; s_\ell'(x)|_{k_{II}r} \}
\nonumber\\
&&- k_{II} \; \Re\{  B^\ell_{F,G} \; e_\ell'(x)|_{k_{II}r}   
e_\ell'(x)|_{k_{II}r} \}.
\end{eqnarray}

\noindent
Finally, for Case II we have the relatively simple result, valid  
for all $r$, that:

\begin{eqnarray}
D^{II}_{\small all~space} &\equiv&
\Im\left\{ {\partial\over\partial r_1} r_1 
           {\partial\over\partial r_2} r_2
\left[ F^{II}_\ell(r_1,r_2) \right]\Big|_{r_1=r_2=r} \right\}
\nonumber\\
&\equiv&
\Im \left\{ {\partial\over\partial r_1} r_1 
            {\partial\over\partial r_2} r_2
\left[ G^{II}_\ell(r_1,r_2) \right]\Big|_{r_1=r_2=r} \right\}
\nonumber\\
&=&
\Im\{ i k_{II} \; (x h_\ell(x))'|_{k_{II}r} \; 
(x  j_\ell(x))'|_{k_{II}r} \}
\nonumber\\
&=&
k_{II} \; \Re\{  e_\ell'(x)|_{k_{II}r} \; s_\ell'(x)|_{k_{II}r} \}
\nonumber\\
&=&
 k_{II} \; \{  [s_\ell'(x)]^2|_{k_{II}r} \}.
\end{eqnarray}

When calculating differences, a large number of pieces in these
expressions quietly cancel.  What remains is still complicated,
but using some identities between Bessel functions and their
derivatives such as

\begin{equation}
{\ell(\ell+1)\over x^2} s_\ell(x) = s_\ell''(x) + s_\ell(x),
\end{equation}

\noindent
and an identical equation that holds for $e_\ell(x)$, the  
calculation simplifies considerably and gives

\begin{eqnarray}
Q_{\small in} &\equiv& \Im\{ X_I(k_{I},r) \}_{\small in}
\nonumber\\
&=&
 {k_{III}\over r^2} \; \sum_{\ell=1}^\infty (2\ell+1) (2 -  
\Re\{A^\ell_F+A^\ell_G\} )
\nonumber\\
&&
\times\left\{ 
2 [s_\ell(x)]^2 + [s_\ell'(x)]^2 + s_\ell(x) s_\ell''(x)  
\right\}|_{k_{III}r}.
\end{eqnarray}

\noindent
Rearranging this yields

\begin{eqnarray}
Q_{\small in} &\equiv& \Im\{ X_I(k_{I},r) \}_{\small in}
\nonumber\\
&=&
2\; {k_{III}\over r^2} \; \sum_{\ell=1}^\infty (2\ell+1)
\nonumber\\
&&\times
\left\{ 
2 [s_\ell(x)]^2 + [s_\ell'(x)]^2  + s_\ell(x) s_\ell''(x)  
\right\}|_{k_{III}r}
\nonumber\\
&-&  {k_{III}\over r^2} \; 
\sum_{\ell=1}^\infty (2\ell+1) \Re\{A^\ell_F+A^\ell_G\} 
\nonumber\\
&&\times
\left\{ 
2 [s_\ell(x)]^2 + [s_\ell'(x)]^2  + s_\ell(x) s_\ell''(x)  
\right\}|_{k_{III}r}.
\nonumber\\
\end{eqnarray}

\noindent
The first two lines are immediately recognizable as the terms we
encountered previously while calculating the energy difference,
$\Delta E = E_{\small Casimir} - E_{\small surface}$.  For these
two terms the sum over $\ell$ can be performed {\em exactly} with
the result that

\begin{eqnarray}
Q_{\small in} &\equiv& \Im\{ X(k_{I},k_{II},r) \}_{\small in}
\nonumber\\
&=&
4 k_{III}^ 3
\nonumber\\
&-&  {k_{III}\over r^2} \; 
\sum_{\ell=1}^\infty (2\ell+1) \Re\{A^\ell_F+A^\ell_G\} 
\nonumber\\
&&\times
\left\{2 [s_\ell(x)]^2 + [s_\ell'(x)]^2  + s_\ell(x) s_\ell''(x)  
\right\}|_{k_{III}r}.
\nonumber\\
\end{eqnarray}

\noindent
Note that for the $A^\ell_{F,G}$ terms we {\em cannot} explicitly
perform the $\ell$ summation because of an implicit $\ell$ dependence
in the coefficients $A^\ell_{F,G}$.  Also notice at this point that
the $A^\ell_{F,G}$--terms in the above expression were the only
pieces retained in~\cite{Milton95,Milton96}.

Taking the cue from the above result, for the correct result in
the region inside the bubble then we can write (using self explanatory
notation)

\begin{eqnarray}
Q_{\small in} &=& 4 k_{III}^ 3 + Q_{\small in}^{\small surface}
\nonumber\\
&=& 4 n_1^3 \; |\omega|^3 + Q_{\small in}^{\small surface}.
\end{eqnarray}

Turning next to the region outside the dielectric sphere, we have

\begin{eqnarray}
Q_{\small out} &\equiv& \Im\{ X_I(k_{I},r) \}_{\small out}
\nonumber\\
&=&
2 {k_{II}\over r^2} \; \sum_{\ell=1}^\infty (2\ell+1)
\nonumber\\
&&\times
\left\{ 2 [s_\ell(x)]^2 + [s_\ell'(x)]^2  + s_\ell(x) s_\ell''(x)  
\right\}|_{k_{II}r}
\nonumber\\
&-&
{k_{II}\over r^2} \; \sum_{\ell=1}^\infty (2\ell+1)
\Re\big[ (B^\ell_F+B^\ell_G)
\nonumber\\
&&\times
\left\{ 2 [e_\ell(x)]^2 + [e_\ell'(x)]^2 + e_\ell(x) e_\ell''(x)  
\right\}
\big]|_{k_{II}r}
\nonumber\\
&=&
4 k_{II}^3 - {k_{II}\over r^2} \; \sum_{\ell=1}^\infty (2\ell+1)
\Re\big[ (B^\ell_F+B^\ell_G)
\nonumber\\
&&\times
\left\{ 2 [e_\ell(x)]^2 + [e_\ell'(x)]^2 + e_\ell(x) e_\ell''(x)  
\right\}
\big]|_{k_{II}r}.
\end{eqnarray}

\noindent
That is

\begin{eqnarray}
Q_{\small out} &=& 4 k_{II}^ 3 + Q_{\small out}^{\small surface}
\nonumber\\
&=& 4 n_2^3 \; |\omega|^3 + Q_{\small out}^{\small surface}.
\end{eqnarray}

Now looking at the Green function for Case II, we get the simple
result that

\begin{eqnarray}
&&\Im\{ X_{II}(k_{II},r) \}_{\small all~space}
= 2{k_{II}\over r^2} \; \sum_{\ell=1}^\infty (2\ell+1)
\nonumber\\
&&\qquad\qquad\times
\left\{ 2 [s_\ell(x)]^2 + [s_\ell'(x)]^2  + s_\ell(x) s_\ell''(x)  
\right\}|_{k_{II}r}
\nonumber\\
&&\qquad\qquad= 4 k_{II}^3 = 4 n_2^3 |\omega|^3.
\end{eqnarray}

Turning again to the total Casimir energy, we can easily perform
the subtractions and substitutions, to obtain

\begin{eqnarray}
\label{E-Casimir-time-split-penultimate}
E_{\small Casimir} &=& 
\int_0^R r^2 dr \int_{0}^{+\infty} {d\omega\over2\pi}
\left[ \cos(\omega\tau) \right] 4 \{k_{III}^3-k_{II}^3\}
\nonumber\\
&&+ \int_0^R r^2 dr \int_{0}^{+\infty} {d\omega\over2\pi}
\left[ \cos(\omega\tau) \right] Q^{\small surface}_{\small in}
\nonumber\\
&&+ \int_R^\infty r^2 dr \int_{0}^{+\infty} {d\omega\over2\pi}
\left[ \cos(\omega\tau) \right] Q^{\small surface}_{\small out}.
\nonumber\\
\end{eqnarray}

\noindent
The remaining $r$ and $\omega$ integrals for the $\omega^3$ term
are easily performed. Reinserting appropriate factors of 
$\hbar$ and $c$ we are led to:

\begin{eqnarray}
E_{\small Casimir} &=& {3\over\pi^2}  \left[ n_1^3 - n_2^3 \right] \hbar c
{1\over  (c \tau)^4} V
\nonumber\\
&&+ \int_0^R r^2 dr \int_{0}^{+\infty} {d\omega\over2\pi}
\left[ \cos(\omega\tau) \right] 
\hbar Q^{\small surface}_{\small in}(k_{III},r)
\nonumber\\
&&+ \int_R^\infty r^2 dr \int_{0}^{+\infty} {d\omega\over2\pi}
\left[ \cos(\omega\tau) \right] 
\hbar Q_{\small out}(k_{II},r) .
\nonumber\\
\end{eqnarray}

\noindent
This is now explicitly of the form:

\begin{center}
(Schwinger's bulk term) + (surface contribution) + $\cdots$
\end{center}

\noindent
In a certain sense this terminates our calculation, since Milton
{\em et al.}\ have already calculated the object we called $E_{\small
surface}$, and explicitly shown it to be a surface term (plus even
higher-order corrections).  For dilute media, $n_1\approx1\approx
n_2$, Milton {\em et al.}\ obtain in eq. (51) of~\cite{Milton95}
and eq. (7.5) of~\cite{Milton96}

\begin{equation}
E_{\small surface} = - (n_1-n_2)^2 \hbar c
\left[ {R^2\over (c\,\tau_{\small physical})^3} + {1\over64R} \right] + \cdots
\end{equation}

\noindent
The existence and qualitative features of this surface
term are in complete agreement with the general analysis adduced in
this paper. 

(See also eq. (51) of~\cite{Milton80} where the
related result for the pressure difference is presented.)

To see what happens for a wave-number cut-off, backtrack to eq.
(\ref{E-Casimir-time-split-intermediate}), insert a wave-number cut-off
$f(k)$ and let the time-splitting parameter go to zero. Then

\begin{eqnarray}
\label{E-Casimir-wave-number-intermediate}
E_{\small Casimir} 
&=&  
\int_0^\infty r^2 dr \; \int_{0}^{+\infty}  {d\omega\over 2\pi}
\nonumber\\
&&
\qquad\times
\; \Im\{f(k_{I}) X_{I}(k_{I}) - f(k_{II}) X_{II}(k_{II},r)\}.
\nonumber\\
&&
\end{eqnarray}

Here again $f(k)$ is any smooth real function with $f(0)=1$ and
$f(+\infty)=0$.  All the computations of the Green functions remain
unaltered, and we replace eq.
(\ref{E-Casimir-time-split-penultimate}) by

\begin{eqnarray}
\label{E-Casimir-wave-number-penultimate}
E_{\small Casimir} &=& 
\int_0^R r^2 dr \int_{0}^{+\infty} {d\omega\over2\pi}
 4 \{f(k_{III}) k_{III}^3- f(k_{II}) k_{II}^3\}
\nonumber\\
&&+ \int_0^R r^2 dr \int_{0}^{+\infty} {d\omega\over2\pi}
\hbar f(k_{III}) Q^{\small surface}_{\small in}
\nonumber\\
&&+ \int_R^\infty r^2 dr \int_{0}^{+\infty} {d\omega\over2\pi}
\hbar f(k_{II}) Q^{\small surface}_{\small out}.
\nonumber\\
\end{eqnarray}

With suitable changes of variable, performing the $r$ integration for
the $k^3$ terms, and reinserting $\hbar$ and $c$ as appropriate,
we have

\begin{eqnarray}
\label{E-Casimir-wave-number-ultimate}
E_{\small Casimir} &=& 
+ 2 V  \int {d^3\vec k\over(2\pi)^3}
 {1\over2}  \hbar f(k) \{\omega_1(k)-\omega_2(k)\}
\nonumber\\
&&+ \int_0^R r^2 dr \int_{0}^{+\infty} {d\omega\over2\pi}
\hbar f(n_1|\omega|) Q^{\small surface}_{\small in}
\nonumber\\
&&+ \int_R^\infty r^2 dr \int_{0}^{+\infty} {d\omega\over2\pi}
\hbar f(n_2|\omega|) Q^{\small surface}_{\small out},
\nonumber\\
\end{eqnarray}

\noindent
which is the central result of this paper.

If we finally approximate air by vacuum, (set $n_1=1$), and make
the above a hard cut-off at wave-number $K$ then the first term is
exactly Schwinger's bulk volume term, while the remaining terms
are subdominant surface and higher-order contributions.

With little additional trouble one can introduce separate wave-number
cut-offs for the two media in which case we can write

\begin{eqnarray}
\label{E-Casimir-wave-number-ultimate-2}
E_{\small Casimir} &=& 
+ 2 V  \int {d^3\vec k\over(2\pi)^3}
 {1\over2}  \hbar \{f_1(k) \omega_1(k)- f_2(k) \omega_2(k)\}
\nonumber\\
&&+ \int_0^R r^2 dr \int_{0}^{+\infty} {d\omega\over2\pi}
\hbar f_1(n_1|\omega|) Q^{\small surface}_{\small in}
\nonumber\\
&&+ \int_R^\infty r^2 dr \int_{0}^{+\infty} {d\omega\over2\pi}
\hbar f_2(n_2|\omega|) Q^{\small surface}_{\small out},
\nonumber\\
\end{eqnarray}

\section{ Discussion}

After this relatively turgid mass of technical manipulations, the
main results of this paper can be succinctly stated:

\noindent
(A) In a dielectric medium of dielectric constant $n$ the Casimir
energy is, in the bulk medium, dominated by a volume term:

\begin{equation}
E^{\small bulk}_{\small Casimir} = {1\over8\pi^2} \; V \; \hbar c \; K^4 \;
\left[{1\over n} - 1 \right].
\end{equation}

\noindent
This result is completely in agreement with Schwinger's argument
in~\cite{Schwinger0}, and can be adduced from simple mode sum
arguments as presented in section \ref{modesum}. We have checked
these mode sum arguments against general field theoretic arguments
in sections \ref{pictures} and \ref{green}.

\noindent
(B) In addition, we pointed out in section \ref{modesum} that
there will be a sub-dominant contribution to the Casimir energy
that is proportional to the surface area of the dielectric. This
surface contribution takes the generic form

\begin{equation}
E^{\small surface}_{\small Casimir} = 
{1\over6\pi^2} \; S \; \hbar c \; K^3 \;  
\left[ {\xi_{\small in }\over n_{\small in }} + 
       {\xi_{\small out}\over n_{\small out}} \right].
\end{equation}

\noindent
This term will be sub-dominant provided the size of the dielectric is 
large compared to the cut-off wavelength. More specifically, provided

\begin{equation}
V/S \gg 1/K = \lambda_0/(2\pi).
\end{equation}

\noindent
(C) In general, following section \ref{modesum}, we can expect
these to be the first two terms of a more general expansion that
includes terms proportional to various geometrical invariants of
the body. By analogy with the situation for non-dispersive Dirichlet,
von Neumann, and Robin boundary conditions~\cite{Balian-Bloch}, we
expect the next term to be proportional to the trace of the extrinsic
curvature integrated over the surface of the body.

\noindent
(D) These very general considerations are buttressed by an explicit
re-assessment, in sections \ref{spherical} and \ref{full}, of
currently extant calculations for a dielectric sphere. We find that
some calculations have used an inappropriate subtraction scheme to
define what is taken to be the Casimir energy.  When this is fixed,
all calculations fall completely into line with the general
considerations presented in this paper.  Furthermore, the corrected
calculations are then also seen to be in complete agreement with
Schwinger~\cite{Schwinger0}.

\noindent
(E) In previous analyses of the Casimir effect it has been common
to fixate on the van der Waal's forces as the underlying physics
ultimately responsible for the Casimir effect.  We strongly disagree
with this point of view, and emphasize, rather, that the presence
of the dielectric medium induces a {\em change} in the dispersion
relation and a {\em change} in the density of states, that results
in a {\em change} in the total zero point energy.  This {\em change}
in the total zero point energy {\em is} the Casimir energy.

For a finite-volume dielectric, relative to vacuum,

\begin{eqnarray}
E_{\small Casimir} &=& 2 V \int {d^3 \vec k\over(2\pi)^3} {1\over2} \hbar
\left[ \omega(k) - c k \right] + \cdots,
\end{eqnarray}

\noindent
where the dots represent terms arising from higher-order distortions 
of the density of states due to finite-volume effects.

\noindent
(F) We close with what is perhaps a trivial point that we nevertheless
feel should be made explicit:  The volume contribution to the
Casimir energy is always there, and always physical, but {\em
sometimes} because of the specific physics of the problem, it is
safe to neglect it.

On the one hand: Suppose we are provided with a fixed number of
dielectric bodies of fixed shape (in particular, of fixed volume),
and suppose that we simply wish to move the bodies around in space
with respect to each other. Then the bulk volume contributions to
the Casimir energy, while still definitely present, are constants
independent of the relative physical location of the dielectric
bodies, and so merely provide a constant offset to the total Casimir
energy. If all we are interested in is the energy differences
between different spatial configurations of the same bodies then
the various volume contributions can be quietly neglected.

On the other hand, an equally physical situation is this: The volume
contribution is of critical importance whenever we want to calculate
the energy difference between an inhomogeneous dielectric and a
homogeneous dielectric wherein the irregularities have been filled
in. {\em This is the physical situation for example in
the case of bubble formation in a dielectric medium.}

\acknowledgements

This work was supported in part by the U.S. Department of Energy,
the U.S. National Science Foundation, by the Spanish Ministry of
Education and Science, and the Spanish Ministry of Defense. Part of
this work was carried out at the Laboratory for Space Astrophysics
and Fundamental Physics (LAEFF, Madrid), and C.E.C., C.M-P., and
M.V. wish to gratefully acknowledge the hospitality shown. Part of
this work was carried out at Los Alamos National Laboratory, and
J.P-M. wishes to gratefully acknowledge the hospitality shown to
him there.

\appendix
\section{Ricatti--Bessel functions}

This appendix collects some useful identities involving sums of
Ricatti--Bessel functions. We start by noting the very useful result
given in Abramowitz and Stegun, page 440, eq.  10.1.45. In
terms of spherical Bessel functions

\begin{equation}
\sum_{\ell=0}^{\infty} (2\ell+1) j_\ell(x) j_\ell(y) =
{\sin(x-y)\over(x-y)}.
\end{equation}

An equivalent result, in terms of the ordinary Bessel functions,
can be obtained by taking the real part of equations 8.533.1
and 8.533.2 on page 980 of Gradshteyn and Ryzhik.

For the purposes of this paper, it is more convenient to use the
Ricatti--Bessel functions defined by

\begin{equation}
s_\ell(x) = x j_\ell(x) = \sqrt{{\pi x\over 2}} J_{{\ell+1/2}}(x).
\end{equation}

\begin{equation}
e_\ell(x) = x h_\ell(x) = \sqrt{{\pi x\over 2}} H_{{\ell+1/2}}(x).
\end{equation}

Furthermore, since the sums occurring in the current problem always
run from $\ell=1$, rather than $\ell=0$, it is useful to pull the
$\ell=0$ terms to the right hand side and thus write

\begin{eqnarray}
\sum_{\ell=1}^{\infty} (2\ell+1) s_\ell(x) s_\ell(y) =
x y {\sin(x-y)\over(x-y)}- \sin(x) \sin(y).
\nonumber\\
\end{eqnarray}

By taking derivatives with respect to $x$ and $y$ it is now
straightforward to show that

\begin{eqnarray}
\sum_{\ell=1}^{\infty} (2\ell+1) s_\ell'(x) s_\ell'(y) &=&
x y {\sin(x-y)\over(x-y)} +  \cos(x-y)
\nonumber\\
&&
+2 x y \left[{\cos(x-y)\over(x-y)^2} -{\sin(x-y)\over(x-y)^3}\right]
\nonumber\\
&&-\cos(x)\cos(y),
\end{eqnarray}
and
\begin{eqnarray}
\sum_{\ell=1}^{\infty} (2\ell+1) s_\ell(x) s_\ell''(y) &=&
-x y {\sin(x-y)\over(x-y)}
\nonumber\\
&&
-2 x^2 \left[{\cos(x-y)\over(x-y)^2}  -{\sin(x-y)\over(x-y)^3}\right]
\nonumber\\
&&
+ \sin(x)\sin(y).
\end{eqnarray}

Taking coincidence limits ($y\to x$) we now get

\begin{equation}
\sum_{\ell=1}^{\infty} (2\ell+1) s_\ell(x) s_\ell(x) =
 x^2 - \sin^2(x),
\end{equation}

\noindent
while

\begin{equation}
\sum_{\ell=1}^{\infty} (2\ell+1) s_\ell'(x) s_\ell'(x) =
{1\over3} x^2 + \sin^2(x),
\end{equation}

\noindent
and

\begin{equation}
\sum_{\ell=1}^{\infty} (2\ell+1) s_\ell(x) s_\ell''(x) =
-{1\over3} x^2 + \sin(x)^2.
\end{equation}

Note in particular that

\begin{equation}
\sum_{\ell=1}^{\infty} (2\ell+1)
\left[ s_\ell(x) s_\ell(x) + s_\ell'(x) s_\ell'(x) \right] 
=
{4\over3} x^2,
\end{equation}

\noindent
and

\begin{eqnarray}
\sum_{\ell=1}^{\infty} (2\ell+1)
\left[ s_\ell'(x) s_\ell'(x) - s_\ell(x) s_\ell''(x) \right]
= 
{2\over3} x^2,
\end{eqnarray}

\noindent
and

\begin{eqnarray}
&&\sum_{\ell=1}^{\infty} (2\ell+1)
\left[ 2 s_\ell(x) s_\ell(x) + s_\ell'(x) s_\ell'(x) + s_\ell(x)  
s_\ell''(x) \right] 
\nonumber\\
&& 
\qquad\qquad = 2 x^2.
\end{eqnarray}

These last two equations are the only sums of Ricatti--Bessel
functions we will actually need. By the manner in which we have
obtained them we can easily see that they are simple generalizations
of the usual summation theorems for ordinary Bessel functions.


\end{document}